\documentclass[11pt]{article}

\usepackage{amsfonts}
\usepackage{amsmath}
\usepackage{amssymb}
\usepackage{stackrel}
\usepackage{setspace}
\usepackage{booktabs}
\usepackage{bm}
\usepackage{booktabs}
\usepackage{geometry}
\usepackage{marvosym}
\usepackage{graphicx}
\usepackage{epstopdf}
\usepackage{lscape}
\usepackage{url}
\usepackage{harvard}

\newtheorem{proposition}{Proposition}

\newcommand{\Exp}{\mathrm{E}}
\newcommand{\Var}{\mathrm{Var}}
\newcommand{\Cov}{\mathrm{Cov}}

\numberwithin{equation}{section}
\allowdisplaybreaks

\input{ee.sty}

\title{\textbf{Modeling and Forecasting Persistent \\ Financial Durations}\footnote{We are grateful to the editor, two anonymous referees, Karim M. Abadir, Laurent E. Calvet, Marcelo Fernandes, Clifford Hurvich, Paolo Zaffaroni and seminar participants at the 18th Symposium of the Society for Nonlinear Dynamics and Econometrics at Novara (April 2010) for many useful comments and suggestions. Special thanks to Liudas Giraitis for invaluable discussions about various aspects of frequency-domain estimation and inference. Jozef Barun\'{\i}k gratefully acknowledges financial support from the Czech Science Foundation under project No. 12-32263S.}}
\author{Filip \v{Z}ike\v{s}\thanks{Corresponding author: Imperial College London, Business School, Exhibition Road, London SW7 2AZ. Phone: +44 79 5791 3476. Email: fzikes@imperial.ac.uk.} \and Jozef Barun\'{\i}k\thanks{Institute of Economic Studies, Charles University, Opletalova 21, 110 00, Prague,  CR and Institute of Information Theory and Automation, Academy of Sciences of the Czech Republic, Pod Vodarenskou Vezi 4, 182 00, Prague, Czech Republic.} \and Nikhil Shenai\thanks{Imperial College London, Business School, Exhibition Road, London SW7 2AZ.}}
\date{\vspace{5mm}This version: \today}

\begin{document}
\maketitle \vspace{-5mm}
\begin{abstract}
\begin{singlespace}
This paper introduces the Markov-Switching Multifractal Duration (MSMD) model by adapting the MSM stochastic volatility model of Calvet and Fisher (2004) to the duration setting. Although the MSMD process is exponential $\beta$-mixing as we show in the paper, it is capable of generating highly persistent autocorrelation. We study analytically and by simulation how this feature of durations generated by the MSMD process propagates to counts and realized volatility. We employ a quasi-maximum likelihood estimator of the MSMD parameters based on the Whittle approximation and establish its strong consistency and asymptotic normality for general MSMD specifications. We show that the Whittle estimation is a computationally simple and fast alternative to maximum likelihood. Finally, we compare the performance of the MSMD model with competing short- and long-memory duration models in an out-of-sample forecasting exercise based on price durations of three major foreign exchange futures contracts. The results of the comparison show that the MSMD and LMSD perform similarly and are superior to the short-memory ACD models.
\end{singlespace}
\end{abstract}
\newpage
\section{Introduction}
Financial durations measure the time elapsed between various financial market events related to transactions arrivals, price fluctuations, or trading volumes. Modeling durations may be useful for measuring and predicting instantaneous volatility and integrated variance and so may aid high-frequency volatility trading and risk management. Exploiting the intimate relationship between durations and volatility, \citeasnoun{ty10} employ parametric duration models to measure daily volatility using high-frequency data. \citeasnoun{ads08} propose a nonparametric duration-based approach to measuring volatility by relying on the properties of Brownian motion. More generally though, durations are useful for gaining more insight into any information events or variables which change values at each tick, as implied by the theory of market microstructure, and thus may be useful for examining a number of interesting economic hypotheses related to trading and price discovery; see \citeasnoun{e00} for an excellent discussion.

A key stylised fact noted in the empirical irregularly-spaced event literature is long memory in financial durations. Ever since the seminal contribution of \citeasnoun{er98}, who introduced the first time-series model for financial durations, a number of studies have documented the slowly decaying autocorrelation function of transaction, price and volume durations; see \citeasnoun{p08} for a detailed literature review. \citeasnoun{dhh10} recently test for long memory in durations and the associated counts and find significant evidence to support the presence of long memory in durations. Despite this empirical regularity, there is currently no paper that explores the alternative models for capturing the persistent autocorrelations of durations and its implications for forecasting. We aim to fill this gap.

Inspired by the success of the Markov-Switching Multifractal (MSM) stochastic volatility model of \citeasnoun{cf04} in forecasting persistent volatility of financial returns, we start by adapting the MSM model to the duration setting, calling the new model the Markov-Switching Multifractal Duration (MSMD) model. This model adds to the class of stochastic durations models of \citeasnoun{bv04} and \citeasnoun{dhh10}, which also evolved from the stochastic volatility literature, though the latent process driving the dynamics of durations in an MSMD is a Markov chain rather than a Gaussian AR(FI)MA process. We show that although the MSMD process is exponential $\beta$-mixing and short-memory, it may exhibit a slowly decaying autocorrelation function over a wide range of lags. This long-memory feature of the process is induced by regime switching of heterogenous persistence: the process is driven by $k$ independent Markov-switching processes with different, though tightly parametrized, transition probabilities.

Relying on the recent results of \citeasnoun{dhsw09} on the propagation of memory from durations to counts and realized volatility, we establish formally that the short memory of MSMD durations translates into short memory in counts and realized volatility. However, within the simple pure-jump model of \citeasnoun{o06}, we show by simulation that despite being a short-memory process, the MSMD model is capable of generating highly persistent realized volatility. Intuitively, the MSMD model lies ``between" the short-memory ACD model of \citeasnoun{er98} and the Long-Memory Stochastic Duration (LMSD) model of \citeasnoun{dhh10} in the sense that its autocorrelation function can decay much slower over a wider range of lags than that of the ACD model, but eventually assumes an exponential rate of decay, unlike the autocorrelation function of the LMSD model.

Second, we propose quasi-maximum likelihood estimation of the MSMD parameters based on the Whittle approximation. The main motivation for exploring this estimation method as an alternative to exact maximum likelihood is computational burden associated with the latter in large samples, and its limitation to the case of an MSM specification with a finite number of states. Contrary to this, the Whittle estimator works in either case and is computationally simple and fast. Relying on results from the statistics literature, we formally establish strong consistency and asymptotic normality of the Whittle estimator under fairly mild assumptions for a wide range of MSMD specifications.

Note that computational speed is not a mere convenience in our context: given the increasing importance of algorithmic and high-frequency traders, who are capable of generating tens of thousands of limit and market orders in a single day, the amount of data usable for estimation has grown enormously in many markets (\citename{hs10}, 2010). For such environments, fast estimation methods simply become a necessity, even with ever-faster modern computers. Last but not least, the Whittle estimator can be easily adapted to the original MSM stochastic volatility model of \citeasnoun{cf04}, and thus represents a contribution to the MSM literature that goes beyond the context of financial durations.

Finally, we compare our estimation and forecasting results with those possible from established duration models. As noted by \citeasnoun{p08}, there is a scarcity of comparisons of duration models, and ideally one would like to undertake a comparison of all the models she has detailed. However, as noted above, only long memory models are able to account for the key stylised fact of long-range dependence in durations. We therefore restrict attention to the LMSD model of \citeasnoun{dhh10}. To investigate the benefits of the relatively complicated MSMD and LMSD models over their simple and easy to estimate short-memory counterparts, we also compare our results with those from the widely used Autoregressive Conditional Duration (ACD) model introduced in the seminal paper by \citeasnoun{er98}. We implement the models on price durations of three major foreign exchange futures contracts traded on the Chicago Mercantile Exchange (CME) in the period between 9 November 2009 and 29 January 2010: Euro, Japanese Yen, and Swiss Franc. We find that while the LMSD and MSMD models deliver generally similar forecast performance, both significantly outperform the ACD model individually, as well as when equally weighted.

The Markov switching multifractal duration model has been proposed independently and in parallel to our work in a recent paper by \citeasnoun{cds12} (henceforth CDS). While the main thrust of CDS is the same - the application of the MSM process of \citeasnoun{cf04} to financial durations - there are several differences that distinguish the two papers. CDS motivate the MSMD model by the mixture-of-distributions hypothesis, whereas our main motivation lies in the long-memory features of the MSM process, and the relationship between persistence of durations and realized volatility. We are not restricting attention to the binomial MSMD model with exponentially distributed innovations, but consider more general versions of the model. Allowing for a wider class of distributions is made possible in practice by employing the Whittle estimator, and it turns out to be empirically beneficial. In terms of empirical application, we differ from CDS by modeling and forecasting price durations as opposed to transactions durations, and focus on foreign exchange futures prices in 2009/2010 rather than individual equities in 1993. Finally, given the high persistence of the durations in our sample, the natural competitor of the MSMD model is the LMSD model rather than the short-memory ACD, and hence, unlike CDS, we include the LMSD model in our forecasting exercise as well.

The rest of the paper is organized as follows. Section 2 introduces the MSMD model and discusses its properties. Section 3 discusses estimation and forecasting for the MSMD model. Section 4 reviews the competing duration models and Section 5 looks at the link between durations, counts and realized volatility. In Section 6 we describe the data and in Section 7 we present estimation and forecasting results. Section 8 concludes. Mathematical proofs are collected in the Appendix.

\section{The MSMD Model}
Let $X_i = t_i - t_{i-1}$ denote the duration between two event arrival times. The three most common events studied in the literature relate to transaction arrivals, price changes and transaction volumes. The Markov-Switching Multifractal Duration (MSMD) model is defined by:
\begin{equation} \label{eq:MSMDx}
X_i =  \psi_i \epsilon_i, \quad \quad i \in \mathbb{Z},
\end{equation}
where $\psi_i$ is the Markov-switching multifractal process of \citeasnoun{cf04};
\begin{equation} \label{eq:MSMDpsi}
\psi_i = \bar{\psi}  \prod_{j=1}^k M_{j,i},
\end{equation}
and $\epsilon_i$ is a sequence of independent unit-mean innovations identically distributed according to some parametric distribution. The latent process in (\ref{eq:MSMDpsi}) is determined by $k$ independent unit-mean multipliers, $M_{j,i}, j=1,...,k$, and a scaling constant, $\bar{\psi}$. At every point in time $i$, each multiplier $M_{j,i}$ takes, with probability $\gamma_j$, a new value $M$ drawn from a common distribution $F_M$, and remains unchanged with probability $1-\gamma_j$:
\begin{equation*}
M_{j,i} = \left\{\begin{array}{ll}
                 M & \text{where $M$ is drawn from $F_M$ with probability } \gamma_j \\
                 M_{j,i-1} & \text{with probability } 1-\gamma_j\\
                 \end{array} \right.
\end{equation*}
The transition probabilities are parsimoniously parametrized by:
\begin{equation}
\gamma_j = 1-(1-\gamma_k)^{(b^{j-k})}, \quad j = 1,...,k,
\end{equation}
where $\gamma_k \in (0,1)$ and $b \in (1,\infty)$. Two specifications for the distribution of the multipliers $F_M$ have been proposed by \citeasnoun{cf04} - binomial and log-normal. In the binomial specification, each multiplier, if at all, is renewed by drawing the values $m_0$ and $2-m_0$; $m_0 \in (1,2)$, with equal probability, ensuring that the mean is equal to one. The transition matrix associated with each multiplier is thus given by:
\begin{equation*}
\mP_j = \left( \begin{array}{cc}
                1-\frac{1}{2}\gamma_j & \frac{1}{2}\gamma_j \\
                \frac{1}{2}\gamma_j & 1-\frac{1}{2}\gamma_j \\
                \end{array} \right).
\end{equation*}
Since the multipliers are independent, the transition matrix of the state vector $\mM_i = (M_{1,i},...,M_{k,i})$ is simply:
\begin{equation} \label{eq:transmat}
\mP = \mP_1 \otimes \mP_2 \otimes \cdots \otimes \mP_k,
\end{equation}
where ``$\otimes$" denotes the Kronecker product. The dimension of the transition matrix is $2^k \times 2^k$ and the state vector takes values in the finite state space $\Omega_{\mM} = \{m_0,2-m_0\}^k$.

In general, any distribution with positive support can be used to model the multipliers. For example, the log-normal specification of \citeasnoun{cf04} replaces the Bernoulli distribution by a log-normal one, i.e. upon switching, the new value of the log multiplier is drawn from $\mathrm{N}(-\lambda,2\lambda)$, where the parameter restriction again imposes unit means for the multipliers. When drawn from a continuous distribution (with respect to Lebesgue measure on $\mathbb{R}_+$), each multiplier assumes a new value with probability one, and the transition kernel of the multiplier is given by
\begin{equation} \label{eq:transkernel}
\mathbb{P}(M_{j,i+h} \in B_j|M_{j,i} = x_j) = (1-(1-\gamma_j)^h)\mathbb{P}(M \in B_j) + (1-\gamma_j)^h \mathbf{1}_{\left\{x_j \in B_j\right\}}, \quad j=1,...,k,
\end{equation}
for any $B_j \in \mathcal{B}(\mathbb{R}_{+})$ and $x_j \in \mathbb{R}_{+}$, where $\mathcal{B}(\mathbb{R}_{+})$ is the Borel $\sigma$-algebra on $\mathbb{R}_{+}$. Since the multipliers are independent, the transition kernel of the chain $\mM_i$ reads
\begin{equation*}
\mathbb{P}(\mM_{i+h} \in B|M_{j,i} = \vx) = \prod_{j=1}^k \mathbb{P}(M_{j,i+h} \in B_j|M_{j,i} = x_j),
\end{equation*}
for any $\vx = (x_1,x_2,...,x_k)'$ and any $B \in \mathcal{B}(\mathbb{R}_{+}^k)$, a Borel $\sigma$-algebra on $\mathbb{R}_{+}^k$, where $B = B_1 \times B_2 \times \cdots \times B_k$, $B_j \in \mathcal{B}(\mathbb{R}_{+})$, $j=1,...,k$. The chain takes values in a state space $\Omega_{\mM} \subseteq \mathbb{R}_{+}^k$.

Having specified the law governing the multipliers, it remains to choose a distribution for the innovations, $\epsilon_i$. As is common in the literature, we consider here the exponential and Weibull distributions. With these specifications of $\epsilon_i$, the law governing the durations, $x_i$, is a mixture of exponentials and a mixture of Weibull distributions, respectively. Imposing a unit mean, the corresponding densities are:
\begin{eqnarray*}
f_E(\epsilon) &=& \exp(-\epsilon) \\
f_W(\epsilon; \kappa) &=& \kappa \xi_W^{\kappa} \epsilon^{\kappa-1} \exp(-\xi_W^{\kappa} \epsilon^{\kappa}), \quad \xi_W = \Gamma(1 + 1/\kappa).
\end{eqnarray*}
For $\kappa = 1$, the Weibull distribution reduces to the exponential distribution with unit mean. Other, more flexible multi-parameter alternatives have been proposed in the context of modeling financial durations: the Burr distribution (\citename{gm00}, 2000) and the generalized gamma distribution (\citename{l99}, 1999), both of which encompass the exponential and Weibull cases. As we are primarily interested in point forecasts in this paper, for the sake of parsimony we confine our attention to the latter two distributions.

\begin{figure}
\centering
\includegraphics[width=\textwidth]{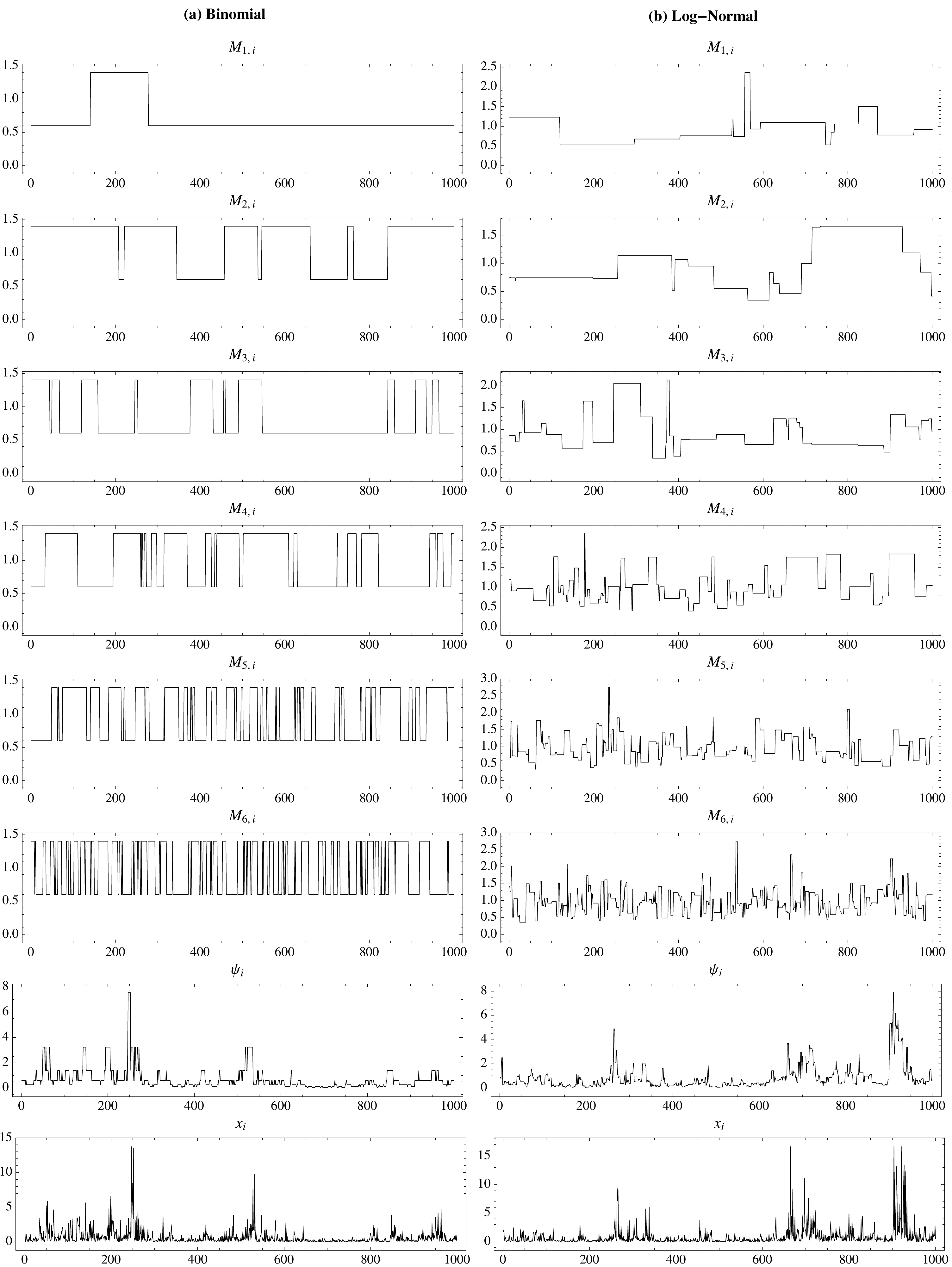}
\caption{Simulated binomial and log-normal MSMD processes with six multipliers and exponentially distributed innovations. The parameters of the processes are $b=3$, $\gamma_k = 0.5$, $m_0 = 1.4$ and $\lambda = 0.15$. \label{fig:msmdsample}}
\end{figure}

To illustrate the behavior of the multipliers and durations in the MSMD model, we plot in Figure \ref{fig:msmdsample} simulated samples from the binomial and log-normal MSMD processes with $k=6$ multipliers and parameters $b=3$, $\gamma_k=0.5$, $m_0 = 1.4$ and $\lambda = 0.15$. In this MSMD specification, $\gamma_1 = 0.0028$ implies that the most persistent multiplier, ($M_{1,i}$), switches, on average, around 3 times in a sample of 1,000 observations if it is drawn from the log-normal distribution, and 1.5 times if it is drawn from the Bernoulli distribution. The least persistent multiplier, ($M_{6,i}$), switches with probability 0.5 and 0.25 in the log-normal and binomial MSMD specifications, respectively. Clearly, both specifications can produce rich dynamics: the duration process is highly persistent but can exhibit sudden erratic movements as observed in empirical data.

\subsection{Stationarity, ergodicity and strong mixing}
It is relatively easy to establish that the Markov chain $\mM_i$ driving the MSMD process is geometrically ergodic as long as the conditions $b > 1$ and $0 < \gamma_k < 1$ are satisfied. Starting with the binomial MSMD specification, we see that under these conditions all elements of the transition matrix of the chain (\ref{eq:transmat}) are strictly positive since $0 < \gamma_j < 1$ for all $j$, and it follows directly from the proof of Theorem 1 in \citename{shiryaev95} (1995, Chapter 1, Section 12) that the chain is geometrically ergodic. The ergodic distribution is given by $\pi_l = 1/2^k$, $l = 1,...,2^k$.

If upon switching the multipliers, $M_{j,i}$, $j=1,...,k$, are drawn randomly from a continuous distribution, $F_M$, with support $\mathbb{R}_+$, the transition kernel associated with the $j$-th multiplier is given in equation (\ref{eq:transkernel}) and the ergodic distribution of the multiplier reads $\pi(B_j):=\lim_{h \rightarrow \infty} \mathbb{P}(M_{j,i+h} \in B_j|M_{j,i} = X_j) = \mathbb{P}(M \in B_j), B_j \in \mathcal{B}(\mathbb{R}_{+})$. Then for any $X_j \in \mathbb{R}_{+}$, $j=1,...,k$ and $h \in \mathbb{N}$,
\begin{equation*}
\sup_{B_j \in \mathcal{B}(\mathbb{R}_{+})}|\mathbb{P}(M_{j,i+h} \in B_j|M_{j,i} = X_j) - \pi(B_j)| \leq (1 - \underline{\gamma})^h,
\end{equation*}
where $0<\underline{\gamma} := \min\{\gamma_1,...,\gamma_k\}<1$. Since the multipliers $M_{j,i}$, $j=1,...,k$, are independent it follows that the chain $\mM_i$ is geometrically ergodic.

Geometric ergodicity of the Markov chain $\mM_i$ in turn implies that the duration process $\{X_i\}$ is strictly stationary $\beta$-mixing with an exponential rate of decay, provided that the chain is initialized from the ergodic distribution. To see this, observe that the duration process belongs to the class of generalized hidden Markov models in the sense of Definition 3 in \citeasnoun{cc02}: the hidden Markov chain $\mM_i$ is strictly stationary and, conditionally on $\mM_i$, the durations $X_i$ are independently distributed where the conditional distribution only depends on $\mM_i$ and not on $i$. Given geometric ergodicity of the hidden chain, Proposition 4 of \citeasnoun{cc02} then implies that the duration process is exponential $\beta$-mixing.

\subsection{Moments, autocovariance function and spectral density}
In Appendix A.1 we show that the first two moments of the MSMD process are given by
\begin{eqnarray}
\Exp(X_i) &=& \bar{\psi}, \label{eq:Meanlevels} \\
\Var(X_i) &=& \bar{\psi}^2[\mathrm{E}(M^2)^k \mathrm{E}(\epsilon_1^2) - 1]. \label{eq:Varlevels}
\end{eqnarray}
The model can exhibit both under- and over-dispersion depending on the distributional assumptions about $M$ and $\epsilon_i$, since the ratio of the variance to the squared mean, $\mathrm{E}(M^2)^k \mathrm{E}(\epsilon_1^2) - 1$, can in general be smaller or larger than one. An MSMD process with exponential innovations, however, always exhibits over-dispersion since for an exponentially distributed $\epsilon_i$, we have $\mathrm{E}(\epsilon_1^2)=2$, and by construction $\mathrm{E}(M^2) > 1$.

An attractive property of the MSMD model is that it possesses a very flexible autocorrelation function (ACF) that can exhibit behaviour similar to long-memory. Appendix A.1 shows that for a general MSMD process with finite $\mathrm{E}(M^2)$ and $\mathrm{E}(\epsilon_1^2)$, we have:
\begin{equation}
\Cov(X_i , X_{i-h}) = \bar{\psi}^2 \biggl(\prod_{j=1}^k [1 + \mathrm{Var}(M)(1 - \gamma_j)^h] - 1 \biggr). \label{eq:ACFlevels}
\end{equation}
and the spectral density is given by
\begin{eqnarray}
f_X(\omega) &=& \frac{1}{2\pi} \bar{\psi}^2 \Exp(M^2)\Var(\epsilon_1) \label{eq:sdX} \\
            && + \frac{\bar{\psi}^2 }{2\pi} \underset{(p_1,...,p_k)\neq (0,...,0)}{\sum_{p_1=0}^1 \sum_{p_2=0}^1 \cdots \sum_{p_k=0}^1}  \left(\frac{\Var(M)^{\sum_{j=1}^k p_j}\left[1-\left(\prod_{j=1}^k (1-\gamma_j)^{p_j}\right)^2\right]}{1 + \left(\prod_{j=1}^k (1-\gamma_j)^{p_j}\right)^2 - 2\left(\prod_{j=1}^k (1-\gamma_j)^{p_j}\right) \cos \omega}\right). \nonumber
\end{eqnarray}
Although (\ref{eq:ACFlevels}) implies that the MSMD process is short-memory as the autocovariance function declines exponentially fast and the spectral density (\ref{eq:sdX}) is bounded at origin, it is capable of mimicking hyperbolic decay over a wide range lags. More specifically, it follows directly from Proposition 1 in \citeasnoun{cf04} that the autocorrelation function of the MSMD durations decays hyperbolically over a large range of lags before transitioning smoothly into exponential decay as the number of multipliers, $k$, grows without bound. Formally, take two arbitrary numbers, $\alpha_1$ and $\alpha_2$ in $(0,1)$, and let $I_k = \{n: \alpha_1 \log_b(b^k) \leq \log_b n \leq \alpha_2 \log_b(b^k)\}$ denote a set of integers containing a wide range of lags. Then
\begin{equation*}
\sup_{n \in I_k} \left|\frac{\log \mathrm{Corr}(X_i,X_{i+n})}{\log n^{-\delta}} - 1 \right| \rightarrow 0
\end{equation*}
as $k \rightarrow \infty$, where $\delta = \log_b(\Exp(M^{2})/[\Exp(M)]^2)$. So despite being a short-memory process, the MSMD model can mimic the persistence of a genuine long-memory process with a hyperbolically decaying autocorrelation function.

For illustration purposes, Figure \ref{fig:theoreticalacf} plots the autocorrelation function of a binomial MSMD process with exponential innovations and various sets of parameter values. We take the case of $k=8$ multipliers and parameters $b=2$, $\gamma_k=0.5$ and $m_0=1.4$, as a benchmark and vary each parameter separately to study how it affects the shape of the autocorrelation function. Increasing $b$ or decreasing $\gamma_k$ both increase the persistence of the process since the switching probabilities of the multipliers decrease (panels (a) and (b)). In the former case, the increase is more pronounced at the long end of the ACF, while in the latter case it affects the short lags of the ACF more. This is due to the different impact of a change in $b$ and $\gamma_k$ on the various switching probabilities as illustrated in panel (a). Increasing the volatility of the multipliers by reducing $m_0$ lowers the multipliers' persistence and thus the persistence of the MSMD process (panel (c)). Finally, increasing the number of multipliers ($k$) while keeping the parameters of the model fixed increases persistence (panel (d)).

\subsection{Exogenous and predetermined variables}
Exogenous or predetermined variables can be easily incorporated into the model by setting $\bar{\psi} = \bar{\psi}_i = \exp(\beta_0 + \vbeta'\vz_i)$, for some vector of variables $\vz_i$. This is useful for several reasons. First, to incorporate the deterministic intraday duration pattern observed in most durations data (\citename{er98}, 1998, \citename{bv04}, 2004, \citename{fg06}, 2006 and \citename{dhh10}, 2010, among many others). Due to the deterministically varying trading activity during the day, the durations tend to be shorter during the early and late trading hours, and relatively longer over lunchtime. Second, one may wish to include additional predictive variables to enhance the forecasting power of the model. A natural candidate when forecasting price durations may be option-implied volatility for which high-frequency data is either available readily (e.g. VIX) or can be constructed from high-frequency options data. Finally, it may be interesting to include some predetermined variables related to market microstructure as in \citeasnoun{er98}, \citeasnoun{bv04} and others.

\section{Estimation, inference and forecasting}
\subsection{Maximum likelihood and optimal forecasting}
The binomial MSM with finite $k$ implies a finite number of states of the hidden Markov process and hence can be estimated by exact maximum likelihood (MLE) via Bayesian updating. This approach has been advocated by \citeasnoun{cf04} for the binomial MSM model of stochastic volatility, and has been shown to work well for sample sizes typically used for estimating models of time-varying volatility. Moreover, the Bayesian filter allows for estimation of the unobserved state probabilities, which in turn permits optimal forecasting. To save space, we omit the details here and refer the reader to \citeasnoun{cf04}.

A disadvantage of the exact maximum likelihood estimator is that it becomes computationally demanding for $k \geq 10$, since the dimension of the transition matrix grows at a rate of $2^k$. Also, it is not applicable to the log-normal MSM process, where the state space of the hidden Markov chain is infinite. These issues have motivated \citeasnoun{l08} to develop a generalized method of moments (GMM) approach, which works for a wide range of MSM specifications and requires only moderate computational resources\footnote{Similarly, \citename{bkm08} (2008, 2012) use a GMM approach to estimate parameters of the Multifractal Random Walk (MRW).}. The drawback of the GMM estimator of \citeasnoun{l08} is that it is applied to the first differences rather than levels of the process and this makes the identification of the parameters $b$ and $\gamma_k$ difficult even when the sample size is very large. \citeasnoun{l08} circumvents this problem by setting these parameters to some pre-specified values that seem to work well for a number of data sets, and estimates by GMM the remaining two parameters only. This may be quite restrictive, however, especially in our context where no previous evidence exists to suggest reasonable values of $b$ and $\gamma_k$ for modeling and forecasting financial durations.

\subsection{Whittle estimation}
We propose an alternative autocovariance-based estimator of the MSMD parameters. In contrast to \citeasnoun{l08} and \citename{bkm08} (2008, 2012) we work in the frequency domain and employ the Whittle quasi-likelihood. An advantage of the Whittle estimation compared to GMM is that it takes into account the entire autocovariance function rather than just a finite subset of lags, and thus avoiding the problem of which autocovariances to match. To obtain better finite-sample properties, we implement the Whittle estimator on logarithmic durations, as the logarithmic durations are much closer to being Gaussian than the durations themselves (see Section 6 for some empirical evidence). Defining $x_i := \log X_i$, $i \in \mathbb{Z}$ and taking logs of both sides of equation (\ref{eq:MSMDx}) we have
\begin{equation*}
x_i =  \log \bar{\psi} + \sum_{j=1}^k m_{j,i} + e_i, \quad i \in \mathbb{Z}, \label{eq:logMSMD}
\end{equation*}
where $m_{j,i} := \log M_{j,i}$ and $e_i := \log \epsilon_i$. We further define $\sigma_m^2 := \Var(m)$ and $\sigma_e^2 := \Var(e_1)$.

It is well-known that for a stationary Gaussian process maximizing the frequency domain representation of the log-likelihood turns out to be asymptotically equivalent to the usual maximum likelihood estimator (\citename{w62}, 1962). The so-called negative Whittle log-likelihood is given by
\begin{equation} \label{eq:WhittleLik}
Q_n(\vtheta) = \frac{1}{n} \sum_{i=1}^{n-1} \left(\log f(\omega_i;\vtheta) + \frac{I_n(\omega_i)}{f(\omega_i;\vtheta)}\right),
\end{equation}
where $f(\omega_i;\vtheta)$ is the spectral density of the logarithmic MSMD process with parameter $\vtheta$, i.e. the spectral density associated to $\{x_j\}$ and
\begin{equation*}
I_n(\omega_i) = \frac{1}{2\pi n}\biggl|\sum_{j=1}^n x_j e^{-\iota \omega_i j}\biggr|^2
\end{equation*}
is the periodogram of the observations $x_1,x_2,...,x_n$, both evaluated at the $i$-th Fourier frequency, $\omega_i = 2 \pi i/n$. The Whittle estimator of $\vtheta$ is obtained by minimizing $Q_n(\vtheta)$:
\begin{equation*}
\hat{\vtheta}_n = \arg \min_{\vtheta \in \Theta} Q_n(\vtheta),
\end{equation*}
Now if the process is not Gaussian, which is our case, minimizing the negative Whittle log-likelihood still works but the resulting estimator is no longer asymptotically equivalent to MLE. The intuition for $\hat{\vtheta}_n$ in the non-Gaussian case is straightforward: under a mixing assumption, the periodogram $I_n(\omega_i)$ is asymptotically distributed as an exponential random variable with parameter $f(\omega_i)$, and for any two Fourier frequencies, $\omega_i$ and $\omega_j$, $i \neq j$, $I_n(\omega_i)$ and $I_n(\omega_j)$ are asymptotically independent (\citename{r73}, 1973). Hence (\ref{eq:WhittleLik}) has a quasi-likelihood interpretation and $\hat{\vtheta}_n $ has been shown to be consistent for $\vtheta$ and asymptotically normally distributed under appropriate regularity conditions.

Implementing the Whittle estimator for the MSMD model is easy since the spectral density is available in closed form. In Appendix A.1 we show that provided the logarithmic MSMD durations possess finite second moments, their autocovariance function reads:
\begin{equation} \label{eq:ACFlogs}
\mathrm{Cov}(x_i, x_{i-h}) = \left \{ \begin{array}{ll}
                                       k\sigma_m^2 + \sigma_e^2 & \mathrm{if} \quad  h=0, \\
                                       \sigma_m^2 \left(\sum_{j=1}^k (1 - \gamma_j)^{|h|}\right) & \mathrm{if} \quad  h \neq 0, \\
                                       \end{array} \right.
\end{equation}
from which the spectral density of the logarithmic MSMD process, $f(\omega)$, can be readily computed via the Fourier transform (see Appendix A.1.). It reads:
\begin{equation} \label{eq:SDlogs}
f(\omega) = \frac{\sigma_m^2}{2\pi} \left(\sum_{j=1}^k \frac{1 - (1 - \gamma_j)^2}{1 + (1 - \gamma_j)^2 - 2(1 - \gamma_j)\cos \omega}\right) + \frac{\sigma_e^2}{2\pi},
\end{equation}
for $\omega \in [-\pi,\pi]$. In the rest of the paper, we will always assume that $\Exp(m^2) < \infty$ and $\Exp(e_1^2) < \infty$ so that (\ref{eq:ACFlogs}) and (\ref{eq:SDlogs}) are well-defined.

We see from equation (\ref{eq:logMSMD}) that the logarithmic MSMD process $x_i$ is a signal-plus-noise process, where the signal is given by a sum of $k$ independent Markov chains and the noise is an $iid$ process independent of the signal. Whittle estimation of signal-plus-noise models has been studied by \citeasnoun{ht82} and \citeasnoun{z09}. Compared to Whittle estimation of linear processes, a complication arises here from the fact that the spectral density of a signal-plus-noise model cannot be be easily factored in the sense that the Whittle log-likelihood cannot be expressed as a sum of two components that depend on disjoint parameter sets. In general, this gives rise to a more complicated limiting distribution of the Whittle estimator.

The asymptotic results obtained by \citeasnoun{ht82} and \citeasnoun{z09} can be applied in our context despite the fact the both the signal and the noise processes have different specifications in these papers. In case of \citeasnoun{ht82}, the signal is an AR(1) process with \emph{iid} innovations, uncorrelated with, thought not necessarily independent of, the \emph{iid} noise process. \citeasnoun{z09} considers a class of models where the signal is an MA($\infty$) process with \emph{iid} innovations and potentially hyporbolically declining MA coefficients (long memory) and allows for correlation between the signal and the \emph{iid} noise. In case of the logarithmic MSMD, while the signal is independent of the noise, it is not a linear process. Given strict stationarity and ergodicity, which was established in Section 2.1, we can, however, invoke the results of \citeasnoun{h73} and establish stong consistency of $\hat{\vtheta}_n$ for an MSMD model.
\begin{proposition}
Let $\{X_i\}$ be an MSMD process with parameter $\vtheta_0 \in \Theta$, where $\Theta$ is a compact subset of the parameter space such that for all $\vtheta_1,\vtheta_2 \in \Theta$, $\vtheta_1 \neq \vtheta_2$ implies $f(\omega;\vtheta_1) \neq f(\omega;\vtheta_2)$ on a set of positive Lebesgue measure. Then $\hat{\vtheta}_n \overset{a.s.}{\longrightarrow} \vtheta_0$ as $n \rightarrow \infty$.
\end{proposition}
The proof is given in section A.2 of the Appendix. The only assumption in Proposition 1 is an identification assumption. Clearly, the Whittle estimator cannot in general work with multi-parameter distributions for the multipliers $M$ and innovations $\epsilon$; it is easy to see from (\ref{eq:SDlogs}) that the Whittle estimator can only identify $\sigma_m^2$ and $\sigma_e^2$. The functions mapping the parameters of the distribution of $M$ and $\epsilon$ into $\sigma_m^2$ and $\sigma_e^2$ have to be continuous, differentiable, one-to-one and onto. This is clearly satisfied for the Bernoulli, log-normal and Weibull distributions. In addition, the Whittle estimator cannot identify the mean of the duration process, $\bar{\psi}$, but this is of lesser concern in our application since durations are typically seasonally pre-adjusted and the model is estimated using the seasonally adjusted durations that have unit mean by construction. Nonetheless, the sample mean can be always used to consistently estimate $\bar{\psi}$ if needed.

Turning to the central limit of $\hat{\vtheta}_n$, we exploit the fact that despite non-linearity, $\{x_i\}$ has a simple vector MA($\infty$) representation (see equation (\ref{eq:marep}) in the Appendix), which allows us to utilize the general results of \citeasnoun{ht82} provided we verify the relevant regularity conditions. This is done in Section A.3 of the Appendix and proves the following.
\begin{proposition}
Let the assumption of Proposition 1 hold with $\Exp(m^{4+r}) < \infty$, $r>0$ and $\Exp(\epsilon_1^4) < \infty$. Then as $n \rightarrow \infty$,
\begin{equation} \label{eq:CLT}
\sqrt{n}(\hat{\vtheta}_n - \vtheta_0) \overset{d}{\longrightarrow} \mathrm{N}(0,\mM(\vtheta_0)^{-1} \mV(\vtheta_0) \mM(\vtheta_0)^{-1}),
\end{equation}
where
\begin{eqnarray}
\mM(\vtheta_0) &=& \frac{1}{2\pi}\int_{-\pi}^{\pi} \left[\vg(\omega;\vtheta)\vg(\omega;\vtheta)'\right]_{\vtheta=\vtheta_0}\mathrm{d}\omega, \label{eq:trueM} \\
\mV(\vtheta_0) &=& \frac{1}{\pi} \int_{-\pi}^{\pi} \left[\vg(\omega;\vtheta)\vg(\omega;\vtheta)'\right]_{\vtheta=\vtheta_0}\mathrm{d}\omega  \label{eq:trueV} \\ && + \frac{1}{2\pi}  \int_{-\pi}^{\pi}  \int_{-\pi}^{\pi} \left[\frac{\vg(\omega_1;\vtheta)}{f(\omega_1;\vtheta)}\frac{\vg(\omega_2;\vtheta)'}{f(\omega_2;\vtheta)} S(-\omega_1,\omega_2,-\omega_2;\vtheta)\right]_{\vtheta=\vtheta_0} \mathrm{d}\omega_1 \mathrm{d}\omega_2, \nonumber
\end{eqnarray}
$\vg(\omega;\vtheta) = \frac{\partial \log f(\omega;\vtheta)}{\partial \vtheta}$ and $S(\omega_1,\omega_2,\omega_3;\vtheta)$ denotes the model trispectrum.
\end{proposition}
The trispectrum entering the limiting variance through (\ref{eq:trueV}) is defined as the Fourier transform of the fourth-order cumulants of $x_i$ (see e.g. \citename{m91}, 1991, for details). It is very difficult to obtain the trispectrum in closed form, except for some special cases. The most simple case arises when both $M$ and $\epsilon_i$ are log-normally distributed, since then $m$ and $e_i$, and thus $x_i$, are Gaussian implying that the fourth-order cumulants of $x_i$ are identically zero and $S(\omega_1,\omega_2,\omega_3;\vtheta) \equiv 0$. The limiting variance of the Whittle estimator in (\ref{eq:CLT}) then reduces to $4\pi \mM(\vtheta_0)^{-1}$.

Relaxing the Gaussianity of $e_i$ while maintaining Gaussianity of $m$ leads to a limiting variance matrix that is no longer robust to fourth-order cumulants, but is still available in closed form. Due to the independence of the multipliers and $e_i$, the cumulants and hence the trispectrum are additive, and since $e_i$ is \emph{iid} with finite fourth moment, the fourth-order cumulants satisfy $\mathrm{cum}(e_i,e_{i+h_1},e_{i+h_2},e_{i+h_3}) = \Exp(e_1^4)$ if $h_1=h_2=h_3=0$, and equal zero otherwise. Thus, $S(\omega_1,\omega_2,\omega_3;\vtheta) = \Exp(e_1^4)/(2\pi)^3$ (e.g. \citename{m91}, 1991). From this point of view, the log-normal specification of the MSM multipliers appears to be particularly attractive in practice, as the limiting variance of the Whittle estimator takes a manageable form and can be easily estimated by the plug-in estimators provided below.

Before we turn to the estimation of the asymptotic variance, we remark that the requirement in Proposition 2 that $4+r$ moment of $m$ exist for some $r>0$, rather than for $r=0$, is dictated precisely by the fact that we are unable to derive the trispectrum in closed form and verify directly that it is well-defined for a general MSMD process. Instead, we have to rely on a mixing inequality to establish that the fourth-order cumulants are absolutely summable, and this requires $r > 0$. Given strict stationarity and exponential strong mixing of $x_i$ we nonetheless conjecture that Proposition 2 holds with $r=0$ as well.

To estimate the asymptotic variance, we can use the following plug-in estimators for $\mM(\vtheta_0)$ and $\mV(\vtheta_0)$:
\begin{eqnarray}
\mM(\hat{\vtheta})&=& \frac{1}{n}\sum_{i=1}^{n-1} \left[\frac{\partial^2f(\omega_i;\vtheta)}{\partial\vtheta \partial\vtheta'}\left(\frac{1}{f(\omega_i;\vtheta)} - \frac{I_n(\omega_i)}{f^2(\omega_i;\vtheta)}\right)\right]_{\vtheta = \hat{\vtheta}}, \label{eq:Mhat} \\
\mV(\hat{\vtheta}) &=& \frac{2}{n}\sum_{i=1}^{n-1} [g(\omega_i;\vtheta)g(\omega_i;\vtheta)']_{\vtheta = \hat{\vtheta}} \label{eq:Vhat} \\
 && + \frac{2\pi}{n^2} \sum_{i_1=1}^{n-1}\sum_{i_2=1}^{n-1} \left[\frac{g(\omega_{i_1};\vtheta)}{f(\omega_{i_1})}\frac{g(\omega_{i_2};\vtheta)'}{f(\omega_{i_2})}S(-\omega_{i_1},\omega_{i_2},-\omega_{i_2};\vtheta)\right]_{\vtheta = \hat{\vtheta}}. \nonumber
\end{eqnarray}
Consistency of $\mM(\hat{\vtheta})$ follows from the consistency of $\hat{\vtheta}_n$ and stochastic equicontinuity of $\mM(\hat{\vtheta})$ where the latter is implied by the smoothness of the third-derivatives of the model spectral density on $\Theta$, see \citeasnoun{z09} for details. For $\mV(\hat{\vtheta})$ consistency cannot be in general established, unless of course one knows the trispectrum in closed form. When this is not the case, we propose a Newey-West estimator:
\begin{eqnarray}
\widehat{\mV(\theta_0)} &=& \frac{1}{n} \sum_{i=1}^{n-1} \left[\frac{\partial q_i(\vtheta)}{\partial \vtheta}\frac{\partial q_i(\vtheta)}{\partial \vtheta'}\right]_{\vtheta = \hat{\vtheta}} \label{eq:VhatNW} \\
&& + \frac{1}{n}\sum_{b=1}^B \sum_{i=b+1}^{n-1}\left(1 - \frac{b}{B+1}\right) \left[\frac{\partial q_i(\vtheta)}{\partial \vtheta}\frac{\partial q_{i-b}(\vtheta)}{\partial \vtheta'} + \frac{\partial q_{i-b}(\vtheta)}{\partial \vtheta}\frac{\partial q_{i}(\vtheta)}{\partial \vtheta'} \right]_{\vtheta = \hat{\vtheta}}, \nonumber
\end{eqnarray}
where $q_i(\vtheta) = \log f(\omega_i;\vtheta) + \frac{I_n(\omega_i)}{f(\omega_i;\vtheta)}$. Alternatively, one can use a similar estimator proposed by \citeasnoun{t82}. A rigorous proof of consistency of $\widehat{\mV(\theta_0)}$ is beyond the scope of this paper and is left for future work.

Before moving onto linear forecasting in the MSMD model, it is interesting to note that the autocovariance function of the logarithmic MSMD process is equivalent to that of a signal-plus-noise process $\{z_i\}$, in which the signal is a sum of $k$ independent AR(1) processes:
\begin{eqnarray*}
z_i &=& \sum_{j=1}^k y_{j,i} + \eta_i, \\
y_{j,i} &=& \rho_j y_{j,i-1} + \eta_{j,i}
\end{eqnarray*}
parametrized by $\rho_j =  1-\gamma_j, \sigma_{\eta_{j,i}}^2 = \sigma_m^2(1 - (1-\gamma_j)^2)$, $j=1,..,k$, and $\sigma_{\eta_i}^2 = \sigma_e^2$. In view of the seminal work of \citeasnoun{g80} on aggregation of short-memory processes of heterogenous persistence, it is hardly surprising to find that as $k \rightarrow \infty$ the MSMD process can generate highly persistent logarithmic durations.

\subsection{Linear forecasting}
When optimal forecasting discussed in the previous subsection is not feasible due to the dimensionality of the state space, \citeasnoun{l08} suggests using best linear forecasts (e.g. \citename{bd91}, 1991). This forecasting rule only requires the knowledge of the autocovariance function of the model and thus works as long as one has a set of consistent parameter estimates at hand, regardless of the estimation method used to obtain them. Formally, an $h$-step ahead forecast based on the most recent $n$ observations, denoted by $\hat{X}_{n+h|n}$, is obtained from
\begin{equation} \label{eq:linearforecasts}
\hat{X}_{n+h|n} = \sum_{j=1}^n \phi_{nj}^{(h)}X_{n+1-j} = \vphi_{n}^{(h)}\mX_n,
\end{equation}
where the vector of weights $\vphi_{n}^{(h)}$ is a solution to $\mGamma_n \vphi_{n}^{(h)} = \vc_n^{(h)}$, in which $\vc_n^{(h)} = (c(h),c(h+1),...,c(n+h-1))'$ denotes the vector of autocovariances of the true process from lag $h$ to lag $n+h-1$, and $\mGamma_n = \{c(i-j)\}_{i,j=1,...,n}$ is the variance-covariance matrix of $\mX_n = (X_1,X_2,...,X_n)'$. The autocovariance function of the MSMD process is provided in (\ref{eq:ACFlevels}) and the weights $\vphi_{n}^{(h)}$ can be efficiently calculated using the generalized Levinson-Durbin algorithm developed by \citeasnoun{bd04}.

\subsection{Specification testing}
To test the goodness of fit of the MSMD model, we employ the specification test of \citeasnoun{chd04}. The idea of the test is to compare the estimated model's spectral density with the smoothed periodogram of the data. Under the null hypothesis of correct model specification, the two should be close. The main advantage of this approach is that the test statistic does not require residuals, which makes it particularly suitable for specification testing of stochastic durations models.

The test statistic is given by
\begin{equation*}
T_n = \left(\frac{2\pi}{n}\sum_{i=0}^{n-1}\tilde{f}(\omega_i)\right)^{-2}\left(\frac{2\pi}{n}\sum_{l=0}^{n-1}\tilde{f}^2(\omega_i)\right),
\end{equation*}
where
\begin{equation*}
\tilde{f}(\omega) = \frac{2\pi}{n}\sum_{i=0}^{n-1}\frac{W(\omega - \omega_i)I_n(\omega_i)}{f(\omega_i;\hat{\vtheta})}, \quad W(\omega) = \frac{1}{2\pi}\sum_{|h|<n}k(h/p_n)e^{-\iota h\omega}
\end{equation*}
$k$ is a symmetric kernel function with $k(0)=1$, and $p_n$ is a bandwidth parameter. Provided that (i) $\hat{\vtheta}$ is $\sqrt{n}$-consistent, (ii) the underlying process $\{x_i\}$ can be written as $x_i = \sum_{l=0}^{\infty} \psi_l \epsilon_{i-l}$, where $\epsilon_{i}$ is \emph{iid} with zero mean, constant variance, and finite eighth moment, and $\sum_{l=0}^{\infty} |\psi_l|l^{1/2} < \infty$, (iii) the model spectral density is bounded away from zero on $[-\pi,\pi]$, (iv) the bandwidth satisfies $\log^6 n/p_n \rightarrow 0$ and $p_n^{3/2}/n \rightarrow 0$, and (v) the kernel satisfies certain regularity considitions, \citeasnoun{chd04} show that:
\begin{equation*}
\frac{n(T_n - C_n(k))}{D_n(k)} \overset{d}{\rightarrow} \mathrm{N}(0,1),
\end{equation*}
where the centering and scaling terms are given by:
\begin{eqnarray*}
C_n(k) &=& \frac{1}{n\pi}\sum_{l=1}^{n-1}(1-l/n)k^2(l/p_n) + \frac{1}{2\pi}, \\
D_n(k) &=& \frac{2}{\pi^2}\sum_{l=1}^{n-2}(1-l/n)(1-(l+1)/n)k^4(l/p_n).
\end{eqnarray*}
The assumptions underlying this result are clearly not satisfied for the logarithmic MSMD as the process is not linear and cannot be written in the form required by (ii) above. The process nonetheless possess a vector MA($\infty$) representation (\ref{eq:marep}) with geometrically declining coefficients and martingale-difference innovations, which leads us to conjecture that the asymptotic normality of the test statistics still holds, thought it remains unclear whether the limiting variance involves fourth-order cumulants. To shed some light on this issue, we examine the distribution of the test statistic $T_n$ for a variety of MSMD specifications by Monte Carlo simulation, leaving the development of a rigorous limit theory for future work. The results are reported at the end of next section.

\subsection{Simulations}
Before taking the model to the data it is worthwhile exploring the finite-sample properties of the maximum likelihood and Whittle estimators. To do that, we run a simple Monte Carlo experiment for the binomial and log-normal MSMD models with $k=8$ multipliers and either exponential or Weibull innovations\footnote{In an earlier version of the paper we also reported MLE simulation results for the MSMD model with Burr and generalized gamma distributions of the innovations. The results are qualitatively similar to the exponential and Weibull cases and show that the ML estimator works well even when the innovations are drawn from multi-parameter distributions.}. Following \citeasnoun{l08} we set the parameters of the MSM process as $b=2$, $\gamma_k = 0.5$, $m_0=1.4$ (binomial) and $\lambda = 0.15$ (log-normal), and the parameter in the Weibull distribution of innovations as $\kappa = 1.45$.

Due to the computational burden associated with the exact maximum likelihood estimator, the number of Monte Carlo replications for MLE is limited to 500, 250, and 100 replications for $n$ = 1000, 2500, and 5000, respectively. All simulation results for the Whittle estimator are based on 1,000 replications, and we also consider very large samples of 10,000 observations, as the application of the Whittle estimator to the MSMD model is new and the large-sample properties have not been investigated by simulation before.

Table \ref{tab:mcsim} summarizes the simulation results. Starting with the maximum likelihood estimator in the binomial MSMD model, we find that MLE delivers accurate and almost unbiased estimates for both exponential and Weibull specifications; the simulated standard errors scale with $\sqrt{n}$ as dictated by asymptotic theory. As expected, the Whittle estimator is less precise than MLE, and it also entails a significant bias in samples smaller than 5,000 observations, particularly for the parameter $b$. The bias, however, disappears in large samples, and the standard errors also scale with $\sqrt{n}$ as claimed in Proposition 2.

Table \ref{tab:mcsimgf} reports the simulated size of the goodness-of-fit test discussed in the previous section. The test is implemented using the Bartlett kernel and setting the bandwidth according to $p_n = 3n^{0.4}$ as in \citeasnoun{chd04}. We use the same MSMD specifications as in the previous simulations and report results for samples of size $n=10,000$. We find that the simulated size is relatively close to the nominal levels across the different MSMD specifications.

\section{Competing duration models}
Models in the duration literature mimic those in the stochastic volatility literature, and might be similarly divided into observable or GARCH-type models and latent factor or Stochastic Volatility (SV)-type models. The Autoregressive Conditional Duration (ACD) model of \citeasnoun{er98} is a member of the former class, and was extended by \citeasnoun{j98} to the Fractionally Integrated ACD (FIACD) model to incorporate long memory. Bauwens and Veredas' \citeyear{bv04} Stochastic Conditional Duration (SCD) model is a latent factor model, and was modified by \citeasnoun{dhh06} to create the LMSD model by letting the latent factor follow a long-memory process.

It is beyond the scope of this paper to review and compare all existing durations models; we refer the reader to a survey by \citeasnoun{p08}. Since we are interested in modeling and forecasting persistent durations, we focus here on those models that can capture slowly decaying autocorrelations. As noted by \citeasnoun{dhh10}, the FIACD model in a not a long-memory model in the usual sense, as it has infinite mean and hence the autocorrelation function does not exist. We are therefore left with the LMSD model as the only genuine long-memory duration model with well-behaved moments. To assess the benefits of using the relatively more complicated MSMD and LMSD models in practice, we also compare their performance with the short-memory ACD model of \citeasnoun{er98}.

\subsection{The ACD model}
Engle and Russell \citeyear{er98} suggest that the durations, $x_i$, obey the following process abbreviated as ACD($p$,$q$):
\begin{eqnarray*}
x_i  &=& \psi _i \varepsilon_i,  \\
\psi_i  &=& \omega  + \sum_{j=1}^q \alpha_j x_{i-j} + \sum_{l=1}^p \beta_l \psi_{i-l}
\end{eqnarray*}
where $\omega$, $\alpha_i$ and $\beta_i$ are parameters to be estimated, $\psi_i$ is the conditional duration, the conditional mean of $x_i$ i.e. $\Exp_{i-1}(x_i) = \psi_i$, and $\varepsilon_i$ is the \emph{iid} duration innovation having a distribution with positive support. Sufficient conditions for positive durations are that $\omega > 0$, $\alpha_j \geq 0$ and $\beta_j \geq 0$. Weak stationarity is guaranteed by $\sum_{j=1}^q \alpha_j + \sum_{j=1}^p \beta_j < 1$. Overall, the model specification is similar to a GARCH model, except that the conditional mean is being modelled as opposed to the conditional volatility. The autocovariance function of the ACD model decays exponentially, thereby not enabling long memory which is signified by hyperbolic decay.

The ACD models can be estimated using maximum likelihood, given the distribution of the disturbance term. \citeasnoun{er98} propose the exponential and Weibull distributions, while \citeasnoun{gm00} suggest the Burr distribution and \citeasnoun{l99} the generalized gamma distributions. An attractive property of the exponential distribution is that the maximum likelihood estimator has a QMLE interpretation, akin to the MLE of GARCH model under normality. Forecasting in the ACD model proceeds via the ARMA representation (see \citename{er98}, 1998 for details).
\subsection{The LMSD model}
\citeasnoun{bv04} propose the the Stochastic Conditional Duration (SCD) model given by:
\begin{eqnarray}
x_i  &=& \varepsilon_i e^{\psi _i},  \\
\psi_i  &=& \omega  + \beta \psi_{i-1}  + u_i, 
\end{eqnarray}
where $\varepsilon_i$ and $u_i$ are mutually independent \emph{iid} innovations and $\omega$ and $\beta$ are parameters to be estimated. Unlike the ACD model, no conditions on parameters are required to ensure positive durations. Also, weak stationarity is guaranteed as long as $\beta$ is less than 1, which is a simpler condition than for the ACD model. Overall, the model specification is similar to a stochastic volatility model.

While the ACD has only one, observable random variable driving the system dynamics, the SCD model has an observable random variable driving the observed duration and a latent random variable, $u_i$, driving the conditional duration (now $e^{\psi_i}$) via an AR(1) process. The extra random variable enables a richer dynamics structure: \citeasnoun{bv04} point out that the parameters governing dispersion ($\sigma$) and persistence ($\beta$) are separated under the SCD model, whereas they are the same in the ACD model ($\alpha + \beta$), so enabling the SCD model to fit a greater variety of persistence-dispersion profiles.

As with the ACD model, the SCD model is only capable of generating geometric decay in the autocovariance function. In order to enable long memory, \citeasnoun{dhh06} introduce the LMSD process, in which the logged conditional duration equation is replaced with:
\begin{equation*}
\psi_i = \omega + \beta\psi_{i-1} + (1-L)^{-d}u_i
\end{equation*}
Here there is more persistence because the logged conditional duration equation has changed from an AR(1) process to an ARFIMA process.

Estimation of the SCD and LMSD models is less straightforward owing to the unobservable factor. \citeasnoun{bv04} advocate employing the Kalman Filter, while \citeasnoun{dhh06} suggest QMLE using the Whittle approximation. We adopt the latter approach here. The Whittle estimator of the parameters is consistent and asymptotically normal. Forecasting the SCD and LMSD models is possible either through calibration of the best linear predictor, as advocated by \citeasnoun{dhh10}, or via the Kalman Filter. While the LMSD process contains an infinite series of coefficients, it is still possible to create a state-space form as observed by \citeasnoun{cp98} and we adopt their approach here.

\section{Relation to counts and realized volatility}
\citeasnoun{dhsw09} and \citeasnoun{dhh10} recently investigate the propagation of memory of durations to counts and thereby realized volatility\footnote{See \citeasnoun{mm08} for a review of the literature on realized volatility.}. They show that if durations have long (short) memory, then under certain conditions the counts have long (short) memory as well. They also note that, alternatively, long memory in realized volatility can be generated by \emph{iid} infinite-variance durations, as originally modeled by \citeasnoun{l00}, where the memory parameter associated to realized volatility is inversely proportional to the tail index of the distribution of durations. These are, of course, two fundamentally different approaches to generating long memory in volatility, and we naturally focus on the former here, not only because the MSMD durations are not \emph{iid}, but also because we find no empirical evidence supporting the infinite variance assumption required by \citeasnoun{l00}.

To fix notation, recall that $t_i$ denotes the time of the $i$-th event, $X_i = t_i - t_{i-1}$ is the duration between two consecutive events, and let $N(t)$ denote the counting process that counts the number of events that have occurred up to time $t$. In more detail, counts and durations are stationary under different measures, since they define the irregularly-spaced event process (a point process) in terms of different sets of events. We refer to these measures as $\mathbb{P}^N$ and $\mathbb{P}$, respectively. As illuminated by \citeasnoun{dhsw09}, the relevant measure depends on how $N(t)$ is calculated: if it is calculated from the opening of the market on a given day, the relevant measure is $\mathbb{P}^N$, while if from the first event on that day, the relevant measure is $\mathbb{P}$. Since most assets tend to be heavily traded after market opening, the difference may be empirically small.

By making use of equivalence theorems (e.g. \citename{n89}, 1989), \citeasnoun{dhsw09} establish conditions under which memory propagates from durations to counts, then to squared returns and realized volatility. In particular, they show that under certain conditions, the short memory of durations generated by a stationary ACD model implies short memory in the associated counts and realized volatility, while the long memory of durations in the LMSD model implies long memory in counts and realized volatility. With respect to the MSMD process now, the following proposition establishes the conditions under which the short-memory feature of the MSMD (for finite $k$) translates into short memory in the induced counts.
\begin{proposition}
Let $\{X_i\}$ be an MSMD process with finite $k$, $\mathrm{E}(M^{3+r}) < \infty$ and $\mathbb{E}(\epsilon_1^{3+r}) < \infty$ for some $r > 0$. Then the induced counting process $N(t)$ satisfies $\mathrm{Var}_N(N(t)) \sim ct$ for some $c < \infty$, where $\mathrm{Var}_N$ denotes the variance under $\mathbb{P}^N$.
\end{proposition}
The proof is provided in Section A.4 of the Appendix. To link the counts and realized volatility, we follow \citeasnoun{dhsw09} and employ the simple continuous-time pure-jump model of \citeasnoun{o06}. The logarithmic price process, $p(t)$, is assumed to have the following dynamics:
\begin{equation} \label{eq:logp}
p(t) = p(0) + \sum_{j=1}^{N(t)} \xi_j, \quad \xi_j \overset{iid}{\sim} \mathrm{N}(0,\sigma_{\xi}^2),
\end{equation}
where $N(t)$ is the counting process defined above and $\xi_j$ is the size of the $j$-th jump, which is assumed to be independent from the counting process. This assumption significantly simplifies the analysis, but it may not be appropriate for all asset classes: a recent study by \citeasnoun{rw11} shows that it is indeed violated in the case of selected individual stocks traded on the New York Stock Exchange.

A natural measure of variation in the model in equation (\ref{eq:logp}) is the quadratic variation given by:
\begin{equation*}
\langle p \rangle_t = \sum_{j=1}^{N(t)} \xi_j^2.
\end{equation*}
The quadratic variation can be estimated consistently by realized variance. Dividing the time interval $[0,t]$ into $n$ non-overlapping intervals of length $\Delta t = t/n$, the realized variance is defined as:
\begin{equation} \label{eq:RV}
RV_{t,n} = \sum_{i=1}^n (p(i\Delta t) - p((i-1)\Delta t))^2.
\end{equation}
It follows from \citeasnoun{dhsw09} that for the MSMD process satisfying the assumptions of Proposition 1, the realized volatility is a short-memory process.

\begin{figure}
\centering
\includegraphics[width=\textwidth]{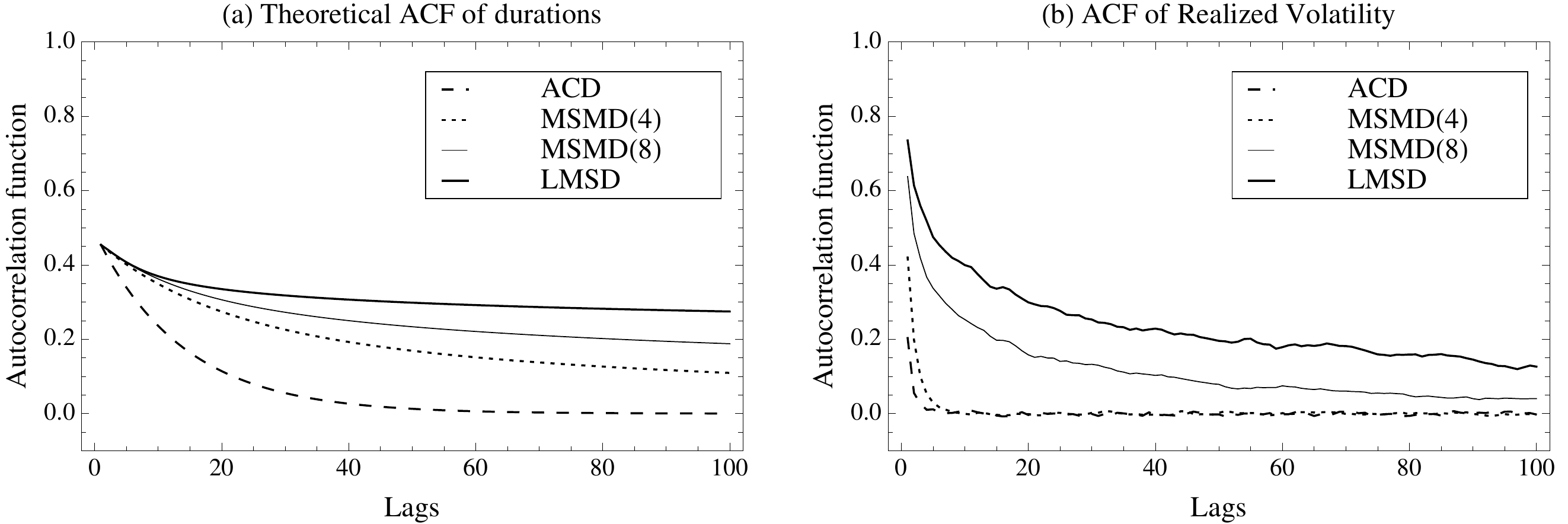}
\caption{(a) Theoretical autocorrelation functions of durations from i) the ACD model with parameters $\alpha=0.24$, $\beta=0.69$, ii) the binomial MSMD(4) model with parameters $m_0=1.84$, $b=3.30$, $\gamma_k=0.047$, iii) the binomial MSMD(8) model with parameters $m_0=1.55$, $b=3.00$, $\gamma_k=0.076$, and iv) the LMSD model with $\omega=1.028$, $\beta=0.73$, $d=0.47$, $\sigma_u^2=0.029$. (b) Simulated autocorrelation functions of daily realized volatility generated by the corresponding duration models i)-iv). \label{fig:acfRV}}
\end{figure}

It is difficult to derive analytically the autocorrelation functions of counts and realized volatility induced by the MSMD process and its competitors. We therefore proceed by simulation. For each duration model, we simulate a trajectory of the induced counting process $N(t)$ and via (\ref{eq:logp}) a trajectory of the associated logarithmic price process $p(t)$. From the simulated price process we then calculate a time series of \emph{daily} realized variance according to (\ref{eq:RV}), where we define one day to have 6.5 hours, or 23,400 seconds. For all duration models, we set the unconditional mean of durations equal to 2 minutes, so that there are around 195 price changes on a typical day in the simulation. The price innovations, $\xi_j$, are drawn randomly from the normal distribution with zero mean and variance $\sigma_{\xi}^2 = 1/195$, implying that the average daily realized variance is around 1\%. Finally, to facilitate comparison we calibrate the parameters of the duration models so that they share the same first-order autocorrelation coefficient, which we set equal to 0.45; see the caption of Figure \ref{fig:acfRV} for the exact parameters values used in the simulations.

Figure \ref{fig:acfRV} plots the theoretical autocorrelation functions of the ACD, MSMD and LMSD durations and their corresponding simulated autocorrelation functions for realized variance as implied by model (\ref{eq:logp}). The figure clearly illustrates how memory propagates from durations to realized volatility. The short-memory ACD model generates realized variance with little persistence, while the long-memory LMSD generates a highly persistent realized variance. The MSMD model is capable of generating both: when the number of multipliers is small ($k=4$), the autocorrelation function of realized variance decays very quickly despite the ACF of durations being quite persistent. Increasing the number of multipliers to 8, the persistence of realized volatility increases dramatically and its ACF now clearly exhibits long-memory features. Thus, despite being short-memory, the MSMD model is capable of generating both highly persistent durations as well as highly persistent realized volatility in the pure jump model (\ref{eq:logp}).

\section{Data Description}
We now apply the MSMD model and its competitors to price durations of three major foreign exchange (FX) futures contracts traded on the Chicago Mercantile Exchange (CME). Our dataset includes all transactions for the Swiss Franc (CHF), Euro (EUR) and Japanese Yen (JPY) futures contracts between 9 November 2009 and 29 January 2010. The data is supplied by TickData, Inc. We focus on the most liquid (front) contracts and restrict attention to the main CME trading hours of 7:20 - 14:00 Chicago time. US and UK Bank holidays are discarded.

Price durations are defined as the minimum time it takes for the price to move by a certain amount. We construct them from the transactions durations, which are simply the durations between successive trades, by a process called thinning. Due to microstructure frictions, such as bid-ask bounce, the price durations may be more informative about the underlying prices process and its volatility as thinning reduces the distortions due to microstructure noise and eliminates duplicate prices, that is transactions with zero price changes. Also, \citeasnoun{er98} show that price durations are closely related to the instantaneous volatility: low price durations imply high instantaneous volatility of the underlying price process, and vice versa.

Correspondingly, we construct the price durations by successively measuring the minimum time required for the futures price to move by at least $c$, starting from the first transaction on each day and discarding overnight durations. The FX futures contracts are highly liquid and usually trade with a tight bid-ask spread of 1-2 ticks, where the tick size equals 0.0001 for CHF, EUR and 0.01 for JPY. To eliminate spurious price changes due to the bid-ask bounce, we set $c = 0.0003$ for CHF and EUR and $c = 0.03$ for JPY. To facilitate comparison across the different currencies, we work with the first 12,000 prices durations for each FX futures contract available in our sample period. The sample size is therefore kept fixed at 12,000 but the sample period varies across the three data sets, though they all start on the 9th November 2009.

Table \ref{tab:descriptivestats} reports the descriptive statistics for the FX futures price durations data. The mean of the price durations is 118s, 106s and 90s for CHF, EUR and JPY, respectively, while the median is around half the mean at 66s, 59s and 43s, respectively, indicating that the distributions of the price durations are heavily positively skewed. The minimum price duration equals 1s for all currencies, while the maximum price duration reaches 42 minutes, 1 hour and 53 minutes, respectively. Consistent with previous empirical evidence, we find that the distribution of price durations exhibits over-dispersion, i.e. the standard deviation of the price durations significantly exceeds the mean by a factor of 1.351, 1.337 and 1.512 for CHF, EUR and JPY, respectively.

It is well-known that the trading activity in most financial markets varies considerably over the course of the day, see e.g. \citeasnoun{er98} who note a hump-shaped pattern for transaction and price durations of individual stocks traded on the New York Stock Exchange (NYSE), with relatively shorter durations at the start and end of the trading day, and longer durations during lunchtime. Consequently, the duration process contains a significant seasonal component that has to be accounted for when estimating a duration model.

There are in principle two ways to do that. First, by incorporating seasonality into the duration models directly and estimating the seasonal parameters jointly with the dynamic parameters of the duration process (\citename{rve07}, 2007). Alternatively, one can first estimate the seasonal component semi- or non-parametrically and fit the duration model to the seasonally-adjusted durations (e.g. \citename{er98}, 1998, and \citename{fg06}, 2006 among many others). \citeasnoun{e00} notes that the large sample sizes typically available in empirical work make the loss of efficiency of the two-step procedure relatively small. Given the complexity of the duration models we are considering in this paper, we opt for the two-stage approach and employ nonparametric regression (the Nadaraya-Watson estimator) to estimate the seasonal component of the price durations, separately for each day of the week as in \citeasnoun{bv04}.

\begin{figure}
\centering
\includegraphics[width=\textwidth]{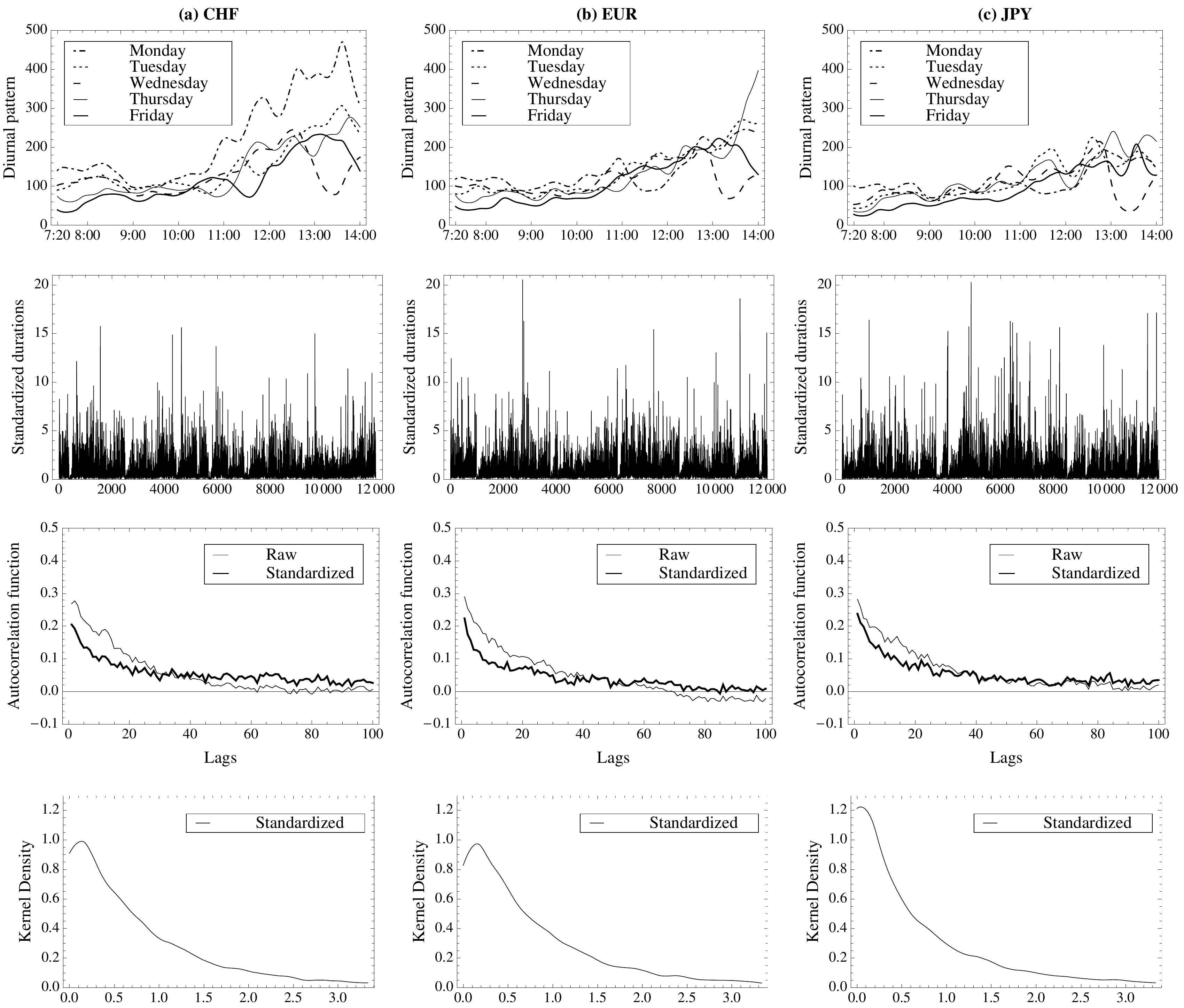}
\caption{Foreign exchange price durations data. The top row shows the diurnal pattern estimated by kernel regression separately for each day of the week. The second row shows the time series of standardized durations and the third row reports the autocorrelation functions of raw and standardized durations. The bottom row plots the empirical density of standardized durations obtained by a boundary-corrected kernel estimator. \label{fig:empirical_data}}
\end{figure}

The estimated intraday seasonal patterns are reported in the top panel of Figure \ref{fig:empirical_data}. The diurnal pattern is relatively stable across the days of the week and currencies up to around 11:00 Chicago time. During this period the U.S. and European trading hours overlap and trading activity in the market is at its peak. After 11:00, trading in London, where a large proportion of global FX trading takes place (\citename{kor11}, 2011), gradually ceases and the average price durations become progressively longer. The exception is Wednesdays, for which we observe a significant dip in the average price durations around 13:30, most likely due to elevated volatility surrounding macroeconomic announcements.

Figure \ref{fig:empirical_data} plots the autocorrelation function of the adjusted durations obtained by dividing the raw durations by the estimated intraday component. Clearly, the persistence in the price durations is not induced by the seasonal component. The descriptive statistics for the adjusted durations are reported in Table \ref{tab:descriptivestats}. The mean is, by construction, close to one, the median remains significantly lower than the mean, and over-dispersion is slightly attenuated by the adjustment. The empirical densities of the standardized durations, estimated in Figure \ref{fig:empirical_data} by a boundary-corrected kernel estimator, are non-monotonic and heavily positively skewed. Finally, we examine the descriptive statistics for the logarithmic standardized durations. We find that the logarithmic durations exhibit negative skewness but almost no excess kurtosis. The tail index estimates obtained by the method of \citeasnoun{hkkp01} indicate that the first 8-9 moments exist, which is in stark contrast to the non-logarithmic durations that only seems to possess the first three moments. The asymptotics for the Whittle estimator discussed in Section 3.2 therefore applies to out logarithmic durations data.

\section{Empirical Results}
The following section compares the estimation and forecasting performance of our MSMD model to the competing ACD and LMSD models. We use the first 10,000 observations for estimation and in-sample specification tests and reserve the remaining 2,000 observations for evaluating out-of-sample forecasting performance. As is common in the durations literature, in the rest of the paper we work exclusively with the seasonally-adjusted durations. Since the mean of the standardized durations is, by construction, close to one, we impose this restriction in all models and do not report the (restricted) estimates of the various constant terms ($\bar{\psi}$ in the MSMD model and $\omega$ in the ACD and LMSD models).

\subsection{Estimation results}
We start by describing the in-sample estimates of MSMD for the three currencies in our sample. We estimate the MSMD model with $k=4$, 6 and 8 multipliers; increasing the number of multipliers beyond 8 does not improve the in-sample and out-of-sample performance of the model. We use exact maximum likelihood to estimate the MSMD model with binomial multipliers and the Whittle estimator for both the binomial and log-normal multipliers. All models are estimated with either the exponential or the Weibull distribution of innovations. Since the MSMD parameter space is not compact ($0 < m_0 < 1$, $\lambda > 0$, $b>1$ and $0<\gamma_k<1$) some constraints are generally required to achieve numerical stability of the optimization routines. For both MLE and Whittle estimation, we use the \verb"MaxSQP" function in the Ox language of \citeasnoun{d06} to maximize the respective objective functions and search over the following parameter space: $m_0 \in [1.001,1.999]$, $\lambda \in [0.001,10]$, $b \in [1.001,10]$ and $\gamma_k \in [0.001,0.999]$.

Tables \ref{tab:MSMDparametersCHF}, \ref{tab:MSMDparametersEUR} and \ref{tab:MSMDparametersJPY} show the estimation results for the CHF, EUR and JPY, respectively. All estimated parameters have reasonable standard errors. The goodness-of-fit test of \citeasnoun{chd04} strongly rejects the null hypothesis of correct model specification for all MSMD models with exponentially distributed innovations. This is generally true for all $k$'s, and across currencies. On the contrary, both the binomial and the log-normal MSMD models with Weibull innovations seem to be correctly specified as we can not reject the null hypothesis at the 5\% level for any of the estimated models. In addition, the log-likelihood is uniformly higher for the binomial MSMD models with Weibull innovations. Thus, the additional flexibility of the Weibull distribution seems to improve the in-sample fit of the MSMD models significantly.

Turning to the number of multipliers, we find that the log-likelihood increases with increasing $k$ in all MSMD models with Weibull innovations. In the case of exponential innovations, the models with six multipliers yield the highest log-likelihood. We have initially experimented with a wider range of values of $k$ and found that going beyond 8 multipliers offers little improvement in terms of both in-sample as well as out-of-sample performance, while reducing $k$ below 4 diminishes performance considerably. The results are available upon request.

\begin{figure}
\centering
\includegraphics[width=\textwidth]{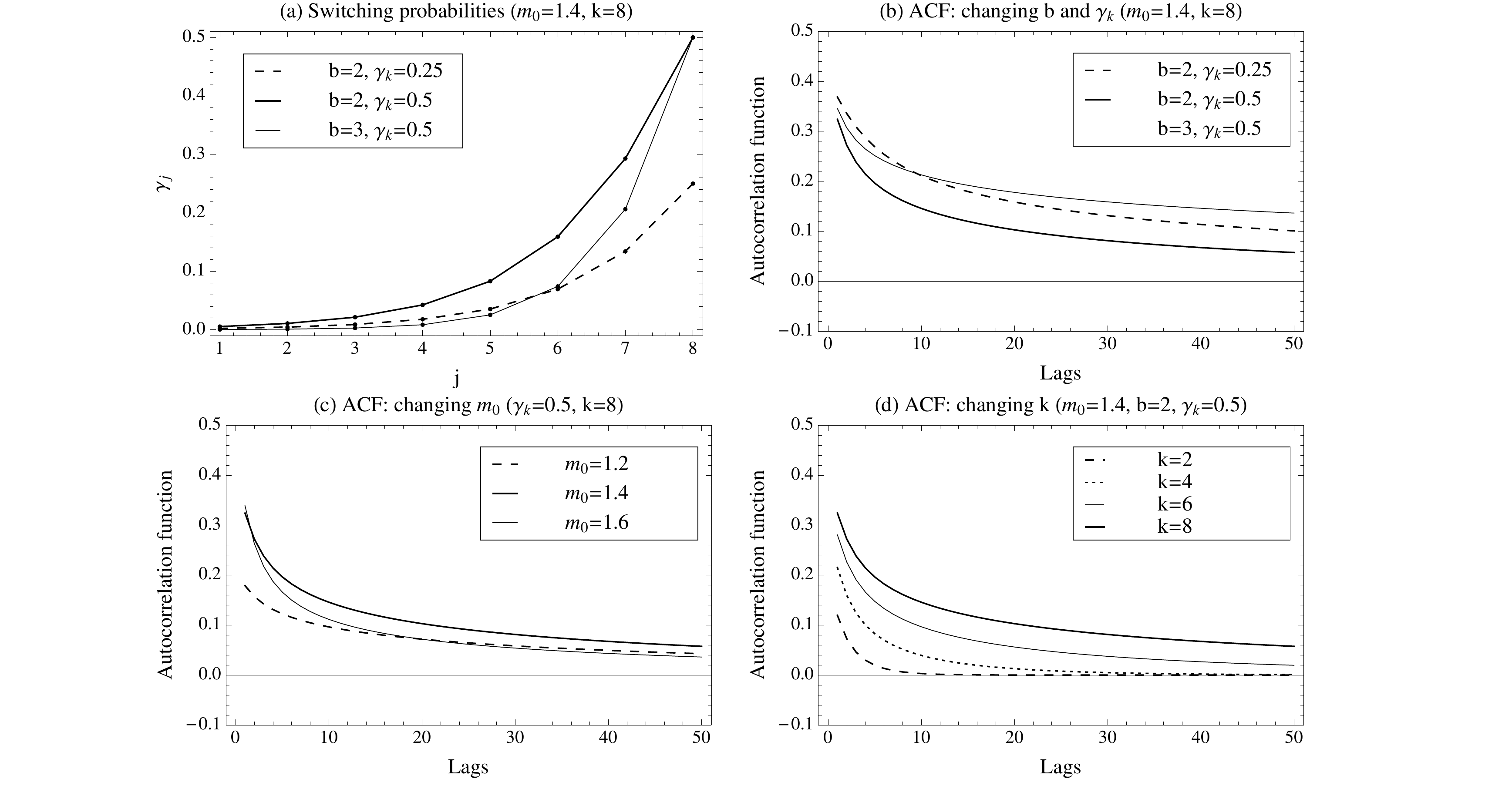}
\caption{Transition probabilities and the autocorrelation function of a binomial MSMD process with exponentially distributed innovations.  \label{fig:theoreticalacf}}
\end{figure}

Comparing the MLE and Whittle parameter estimates for the binomial MSMD specifications, we find that the latter are typically smaller than the former, but generally exhibit a similar pattern. Specifically, both the MLE and Whittle estimates of $b$ tend to decrease with increasing $k$, while the estimates of $\gamma_k$ tend to increase. Intuitively, holding all parameters fixed, increasing the number of multipliers increases the persistence of the MSMD process (see Figure \ref{fig:theoreticalacf}(d)), and hence to fit a given persistence in the data the parameters $b$ and $\gamma_k$ must fall and/or rise, respectively, to compensate (see Figure \ref{fig:theoreticalacf} (b)). Additionally, we observe that the estimates of $m_0$ fall with increasing $k$, in order to compensate for the increase in unconditional variance of the MSMD process associated with rising $k$ (see equation (\ref{eq:Varlevels})). A similar pattern is found for the parameter $\lambda$ in the specification with log-normal multipliers. Note that it is not surprising to find that the Whittle estimates of $b$, $\gamma_k$ and $\kappa$ (when applicable) are the same across the binomial and log-normal specifications; the two model spectral densities only differ in the parametrization of $\mathrm{Var}(\log M)$, see equation (\ref{eq:SDlogs}). This does not, however, imply that the linear forecasts of durations obtained from these models will be the same. The linear forecasts of durations depend on $\mathrm{E}(M^2)$ and $\mathrm{Var}(M)$ (see equations (\ref{eq:Varlevels}), (\ref{eq:ACFlevels}) and (\ref{eq:linearforecasts})), and the fact that $\mathrm{Var}(\log M)$ is the same across the binomial and log-normal specifications does not imply that $\mathrm{E}(M^2)$ and $\mathrm{Var}(M)$ are as well. This will generally be the case whenever the parameter estimates are obtained by implementing the Whittle estimator on non-linearly transformed durations (logs in the present application), rather than the durations themselves.

Having estimated the MSMD model, we now turn to the competing duration models. Table \ref{tab:ACDLMSDparameters} shows the results from estimating the exponential and Weibull ACD and LMSD models for the three FX futures price durations. All estimated parameters have reasonable standard errors. The ACD model with Weibull innovations achieves higher log-likelihood than the ACD model with exponentially distributed innovations, but none of these models generate higher log-likelihoods than the corresponding binomial MSMD models estimated by maximum likelihood. The ACD parameter estimates are qualitatively similar (relatively high $\beta$ and small $\alpha$) and imply very high persistence as $(\alpha + \beta)$ is close to one. High persistence is also implied by the LMSD parameter estimates, where the long memory parameter estimates ($d$) lie between 0.37 and 0.50. It is difficult to assess the relative in-sample fit of the Weibull and exponential LMSD specifications, since the Whittle quasi-likelihoods are not directly comparable.

\subsection{Out-of-sample forecasting performance}
Our main interest lies in relative forecasting performance rather than in the in-sample fit of the various duration models. As we experiment with alternative estimation methods (MLE vs. Whittle) and forecasting schemes (optimal vs. linear), we are really going to be comparing alternative forecasting methods rather than models (\citename{gw06}, 2006). The goal is to shed light not only on the relative ability of the alternative models to capture persistence in the data, but also on the impact of parameter uncertainty and the choice of forecasting rule on relative predictive performance. Specifically, we compare the following methods: (a) optimal forecasts from binomial MSMD(6) or MSMD(8) models estimated by maximum likelihood; (b) linear forecasts from binomial and log-normal MSMD(6) or MSMD(8) models estimated by the Whittle estimator; (c) Kalman filter-based forecasts from the LMSD model estimated by the Whittle estimator; and (d) ARMA representation-based forecasts from the ACD model estimated by maximum likelihood. We also experiment with equally-weighted combinations of (a) and (c), and (b) and (c), as model averaging may help reduce model uncertainty.

We compute and evaluate one step ahead and cumulative 5, 10 and 20 step ahead forecasts of price durations. The cumulative $h$-step ahead forecast, which we denote by $x_{n,h}$, are obtained from the usual multi-step ahead forecast by $x_{n,h} = \sum_{j=1}^h x_{n+j|n}$. Thus, $x_{n,h}$ forecasts the time it takes for $h$ price changes  to occur, as opposed to $x_{n+h|n}$, which forecasts the time elapsed between the $(h-1)$-th and $h$-th price changes. We focus on the cumulative forecasts as they are more interesting in applications, for example in predicting realized variance. We evaluate the accuracy of the forecasts using two common loss functions, the mean square error (MSE) and the mean absolute deviation (MAD), and assess the differences between models statistically by the \citeasnoun{dm95} test for equality of forecast accuracy; the Newey-West estimator is used in the denominator of the Diebold-Mariano test statistic to account for autocorrelation in the multi-step forecasts. Our benchmark against which we assess the MSMD and LMSD models is the short-memory ACD, and we compare models with exponential and Weibull innovations separately.

Tables \ref{tab:forecastsMSE} and \ref{tab:forecastsMAD} report the results of the forecasting performance of the different methodologies. Although there is no uniform ranking across the currencies, forecast horizons and loss functions, a few clear patterns emerge from the exercise. Both the LMSD and MSMD forecasts generally outperform the ACD forecasts in terms of both the MSE and MAD. The gains in forecasting performance increase with the forecast horizon and are generally statistically significant at the 5\% level. The MSMD model performs better when the parameters are estimated by maximum likelihood and the optimal forecasting rule is used, but the linear forecasting scheme coupled with parameter estimates obtained by Whittle estimation also deliver better performance than the ACD, although the difference is not always statistically significant. The superior in-sample fit of the models with Weibull innovations that we documented in the previous section does not necessarily translate into better our-of-sample performance. Similarly, while the MSMD(8) has a higher log-likelihood in-sample, it does not always outperform the MSMD(6) specification.

The LMSD and MSMD models generally perform on par if optimal forecasting and MLE estimates are used for the latter model, with the MSMD sometimes producing slightly better results. The forecast combinations of the LMSD and MSMD models almost always significantly outperform the ACD model, and this is generally true regardless of the estimation method and forecasting rule used for the MSMD model. This is a potentially important result for practitioners, for the Whittle estimators of both MSMD and LMSD parameters are very easy to implement regardless of the size of the sample or the various distributional assumptions made. Thus, we conclude that the long-memory duration models do provide better forecasts than the simple short-memory ACD model.

\section{Conclusion}
This paper introduces a new model for financial durations, featuring persistence that translates from durations to realized volatility. We establish the main properties of the model and propose the Whittle estimator of its parameters as an alternative to maximum likelihood. The asymptotics we obtain for the estimator is by no means confined to the MSMD specifications explored in this paper, and can be readily adapted to other MSM applications, such as stochastic volatility modeling. In an empirical application, we show that the MSMD model performs well in multi-step forecasting.

There are several avenues for future research. It would be worthwhile to experiment with the ``enhanced'' Whittle estimator proposed by \citeasnoun{dhl06} in order to improve the finite-sample properties. The idea of this approach is to apply the Whittle estimator to durations transformed as $\frac{1}{v}X^v$, $v>0$, rather than to logarithmic durations as we did in this paper, as the distribution of $\frac{1}{v}X^v$ may be closer to Gaussian for some $v$ than the distribution of $\log X$. \citeasnoun{at05} show that this transformation is smooth in the sense that the autocovariance function of $\frac{1}{v}X^v$ approaches the autocovariance function of $\log X$ as $v \rightarrow 0$. The moments and spectral density of $\frac{1}{v}X^v$ can be obtained in closed form by following the same steps as in Appendix A.1, which facilitates implementation.

On the empirical side, it would be interesting to use the MSMD model in various risk-management applications. Given the success of the model in multi-step forecasting, one may for example explore its ability to forecast realized volatility over short time-horizons, such as 1 hour, and compare the resulting forecasts with those obtained from popular time-series models for realized volatility. Similarly, the model may be fit to volume durations, and used to predict market trading activity with the aim to optimally time trade execution. We will explore these applications in future work.

\newpage
\bibliographystyle{dcu}
\bibliography{msmd}

\clearpage
\appendix
\footnotesize
\begin{singlespace}
\section{Mathematical appendix}
\subsection{Derivation of autocovariance functions in (\ref{eq:ACFlevels}) and (\ref{eq:ACFlogs})}
Defining $m_{j,i} = g(M_{j,i})$ for some function $g:\mathbb{R} \rightarrow \mathbb{R}$ such that $\mathbb{E}(g^2(M)) < \infty$, we start by showing that:
\begin{equation} \label{eq:Amm}
\mathrm{E}(m_{j,i}m_{j,i-h}) = \mathbb{E}(m_{j,i})^2 + \mathrm{Var}(m_{j,i})(1-\gamma_j)^h.
\end{equation}
for $h \geq 0$. In the binomial MSMD model, the multiplier $M_{j,i}$, if it switches, takes the value of $m_0$ or $(2-m_0)$ with equal probability. To simplify notation, define $p_j := 1 - \frac{1}{2}\gamma_j$, $m_{0,1} := g(m_0)$, $m_{0,2}:= g(2-m_0)$, $\vm_0 := (m_{0,1}, m_{0,2})'$. Then the transition matrix, $\mP_j$, associated with the $j$-th multiplier can be written as:
\begin{equation*}
\mP_j = \left( \begin{array}{cc}
                p_j & 1-p_j \\
                1-p_j & p_j \\
                \end{array} \right) = \left( \begin{array}{cc}
                \frac{\sqrt{2}}{2} & \frac{\sqrt{2}}{2} \\
                \frac{\sqrt{2}}{2} & -\frac{\sqrt{2}}{2} \\
                \end{array} \right) \left( \begin{array}{cc}
                1 & 0 \\
                0 & 2(p_j-1) \\
                \end{array} \right)\left( \begin{array}{cc}
                \frac{\sqrt{2}}{2} & \frac{\sqrt{2}}{2} \\
                \frac{\sqrt{2}}{2} & -\frac{\sqrt{2}}{2} \\
                \end{array} \right) = \mC \mA_j \mC',
\end{equation*}
where $\mC$ is the matrix of eigenvectors of the transition matrix and $\mA_j$ holds the corresponding eigenvalues. Then by the Law of Iterated Expectations (LIE),
\begin{eqnarray*}
\mathrm{E}(m_{j,i}m_{j,i-h}) &=& \mathrm{E}(m_{j,i}m_{j,i-h}|m_{j,i-h}=m_{0,1})\mathbb{P}(m_{j,i-h}=m_{0,1}) \\
                             && \qquad \qquad + \quad \mathrm{E}(m_{j,i}m_{j,i-h}|m_{j,i-h}=m_{0,2})\mathbb{P}(m_{j,i-h}=m_{0,2}), \\
                             &=& \frac{1}{2}m_{0,1}\mathrm{E}(m_{j,i}|m_{j,i-h}=m_{0,1}) + \frac{1}{2}m_{0,2}\mathrm{E}(m_{j,i}|m_{j,i-h}=m_{0,2}), \\
                             &=& \frac{1}{2}\vm_0'\mP_j^h \vm_0, \\
                             &=& \frac{1}{2}\vm_0'\mC \mA_j^h \mC' \vm_0, \\
                             &=& \frac{1}{4}(m_{0,1} + m_{0,2})^2 + \frac{1}{4}(1-\gamma_j)^h(m_{0,1} - m_{0,2})^2, \\
                             &=& \mathbb{E}(m_{j,i})^2 + \mathrm{Var}(m_{j,i})(1-\gamma_j)^h.
\end{eqnarray*}
When the multiplier $M_{j,i}$ is drawn from a continuous distribution upon switching, then the new value it takes is different from the current value with probability one. Then we have:
\begin{eqnarray*}
\mathrm{E}(m_{j,i}m_{j,i-h}) &=& \mathbb{E}[\mathbb{E}(m_{j,i}m_{j,i-h}|m_{j,i-h}), \\
                             &=& \mathbb{E}[m_{j,i-h} \mathbb{E}(m_{j,i}|m_{j,i} \neq m_{j,i-h})\mathbb{P}(m_{j,i} \neq m_{j,i-h}) \\
                             && \qquad \qquad + \quad m_{j,i-h}\mathbb{E}(m_{j,i}|m_{j,i} = m_{j,i-h})\mathbb{P}(m_{j,i} = m_{j,i-h})]\\
                             &=& \mathbb{E}[m_{j,i-h}(\mathbb{E}(m_{j,i})(1 - (1-\gamma_j)^h) + m_{j,i-h}^2 (1-\gamma_j)^h)] \\
                             &=& \mathbb{E}(m_{j,i})^2(1 - (1-\gamma_j)^h) + \mathbb{E}(m_{j,i}^2)(1-\gamma_j)^h, \\
                             &=& \mathbb{E}(m_{j,i})^2 + \mathrm{Var}(m_{j,i})(1-\gamma_j)^h.
\end{eqnarray*}
as claimed.

Now given that the multipliers and $\epsilon_i$ are all mutually independent, we obtain by LIE and (\ref{eq:Amm}) for $h > 0$:
\begin{eqnarray*}
\mathrm{Cov}\left(X_i,X_{i-h} \right) &=&  \mathrm{Cov} (\psi_i \epsilon_i,\psi_{i-h} \epsilon_{i-h}), \\
                                                                   &=&  \mathrm{E}(\psi_i \psi_{i-h})\mathrm{E}(\epsilon_i)\mathrm{E}(\epsilon_{i-h}) - \mathrm{E}(\psi_i) \mathrm{E}( \psi_{i-h})\mathrm{E}(\epsilon_i)\mathrm{E}(\epsilon_{i-h}), \\
                                                                   &=& \bar{\psi}^2 \left[\prod_{j=1}^{k} \mathrm{E} (M_{j,i} M_{j,i-h}) - \left(\prod_{j=1}^{k}\mathrm{E}(M_{j,i})\right)^2\right], \\
                                                                   &=& \bar{\psi}^2 \biggl(\prod_{j=1}^k [1 + \mathrm{Var}(M)(1 - \gamma_j)^h] - 1 \biggr), \nonumber
\end{eqnarray*}
and for $h=0$:
\begin{eqnarray*}
\mathrm{Var}\left(X_i \right) &=& \mathrm{E}(\psi_i^2) \mathrm{E}(\epsilon_i^2) - \mathrm{E}(\psi_i)^2 \mathrm{E}(\epsilon_i)^2, \\
                                          &=& \bar{\psi}^2[\mathrm{E}(M^2)^k \mathrm{E}(\epsilon_i^2) - 1] \nonumber
\end{eqnarray*}
as claimed. Turning to the spectral density, take the discrete Fourier transform of the autocovariance function:
\begin{eqnarray}
\frac{2\pi}{\bar{\psi}^2} f_X(\omega) &=& \sum_{h=-\infty}^{\infty} \frac{1}{\bar{\psi}^2} \Cov(X_i,X_{i-|h|})e^{-\imath \omega h}, \nonumber \\
                           &=& \Var(X_i) - [(1+\Var(M))^k - 1] + \sum_{h=-\infty}^{\infty}  \biggl( \prod_{j=1}^k (1 + \Var(M) (1-\gamma_j)^{|h|})  - 1 \biggr)e^{-\imath \omega h},   \nonumber \\
                           &=& \Exp(M^2)\Var(\epsilon_1) + \sum_{h=-\infty}^{\infty} \underset{(p_1,...,p_k)\neq (0,...,0)}{\sum_{p_1=0}^1 \sum_{p_2=0}^1 \cdots \sum_{p_k=0}^1} \biggl(\Var(M)^{\sum_{j=1}^k p_j} \prod_{j=1}^k (1-\gamma_j)^{|h|p_j}\biggr)e^{-\imath \omega h}, \nonumber \\
                           &=& \Exp(M^2)\Var(\epsilon_1) +  \underset{(p_1,...,p_k)\neq (0,...,0)}{\sum_{p_1=0}^1 \sum_{p_2=0}^1 \cdots \sum_{p_k=0}^1} \left(\Var(M)^{\sum_{j=1}^k p_j}\right) \sum_{h=-\infty}^{\infty} \biggl(\prod_{j=1}^k (1-\gamma_j)^{p_j}\biggr)^{|h|} e^{-\imath \omega h} \nonumber \\
                           &=& \Exp(M^2)\Var(\epsilon_1)\label{eq:sdXl} \\
                           && + \underset{(p_1,...,p_k)\neq (0,...,0)}{\sum_{p_1=0}^1 \sum_{p_2=0}^1 \cdots \sum_{p_k=0}^1}  \left(\frac{\Var(M)^{\sum_{j=1}^k p_j}\left[1-\left(\prod_{j=1}^k (1-\gamma_j)^{p_j}\right)^2\right]}{1 + \left(\prod_{j=1}^k (1-\gamma_j)^{p_j}\right)^2 - 2\left(\prod_{j=1}^k (1-\gamma_j)^{p_j}\right) \cos \omega}\right), \nonumber
\end{eqnarray}
where we use the multi-binomial theorem and the well-known fact that for any $\rho \in (-1,1)$,
\begin{equation} \label{eq:sdar1}
\sum_{h=-\infty}^{\infty}\rho^{|h|}e^{-\imath \omega h} = \frac{1-\rho^2}{1 + \rho^2 - 2\rho \cos \omega},
\end{equation}
since $(p_1,...,p_k) \neq (0,...,0)$ implies that $|\prod_{j=1}^k \delta_j^{p_j}| < 1$.

Turning to the autocovariance function for logarithmic durations, given that the multipliers and $\epsilon_i$ are all independent, we obtain by (\ref{eq:Amm}) for $h \neq 0$:
\begin{equation*}
\mathrm{Cov}(x_i,x_{i-h}) = \sum_{j=1}^k \mathrm{Cov}(m_{j,i},m_{j,i-h}) = \mathrm{Var}(\log M) \sum_{j=1}^k (1-\gamma_j)^h,
\end{equation*}
and for $h=0$:
\begin{equation*}
\mathrm{Var}(x_i) =  \sum_{j=1}^k \mathrm{Var}(m_{j,i}) + \mathrm{Var}(\log \epsilon_i) = k\mathrm{Var}(\log M) + \mathrm{Var}(\log \epsilon_i),
\end{equation*}
as claimed. The spectral density then follows directly by calculating the discrete Fourier transform of the autocovariance function:
\begin{eqnarray*}
2\pi f(\omega) &=& \sum_{h=-\infty}^{\infty} \mathrm{Cov}(x_i, x_{i-|h|})e^{-\imath \omega h}, \\
            &=& \mathrm{Var}(\log \epsilon_1) + \sum_{h=-\infty}^{\infty} \left(\mathrm{Var}(\log M) \sum_{j=1}^k (1-\gamma_j)^{|h|} \right) e^{-\imath \omega h}, \\
            &=& \mathrm{Var}(\log \epsilon_1) + \mathrm{Var}(\log M) \sum_{j=1}^k \frac{1 - (1 - \gamma_j)^2}{1 + (1 - \gamma_j)^2 - 2(1 - \gamma_j)\cos \omega}.
\end{eqnarray*}

\subsection{Proof of Proposition 1}
We follow \citename{z09} (2009, Theorem 1) and rely on the proof of \citename{h73} (1973, Lemma 1). In particular, we show that $\sup_{\vtheta \in \Theta}|Q_n(\vtheta) - Q(\vtheta)| \overset{a.s.} \longrightarrow 0$ as $n \rightarrow \infty$, where $Q(\vtheta) = \frac{1}{2\pi}\int_{-\pi}^{\pi} \log(f(\omega;\vtheta))\mathrm{d}\omega + \frac{1}{2\pi}\int_{-\pi}^{\pi}f(\omega;\vtheta_0)f^{-1}(\omega;\vtheta)\mathrm{d}\omega$, and that $Q(\vtheta) \geq Q(\vtheta_0)$ for any $\vtheta \in \Theta$ with equality holding only if $\vtheta = \vtheta_0$. The statement in the proposition then follows.

Starting with $\frac{1}{n}\sum_{i=1}^{n-1} I_n(\omega_i)/f(\omega_i;\vtheta)$, observe that the continuity and boundedness away from zero of $f(\omega;\vtheta)$ on $[-\pi,\pi]$ implies that the Cesaro sum of the Fourier series of $f^{-1}(\omega;\vtheta)$ taken to $M$ terms converges to $f^{-1}(\omega;\vtheta)$ uniformly on $[-\pi,\pi] \times \Theta$ as $M \rightarrow \infty$ (e.g. \citename{bd91}, 1991, Theorem 2.11.1). Thus the uniform convergence a.s. of $\frac{1}{n}\sum_{i=1}^{n-1} I_n(\omega_i)/f(\omega_i;\vtheta)$ follows by the same argument as in Hannan (1993, Lemma 1), provided that (a) $\frac{1}{n} \sum_{i=h+1}^{n-1} x_i x_{i-h} \overset{a.s.}{\longrightarrow} \Exp(x_i x_{i-h})$ as $n \rightarrow \infty$ for all $h$, $0 \leq h \leq M$, and (b) $x_{j}x_{n-m+j}/n \overset{a.s.} \longrightarrow 0$ as $n \rightarrow \infty$ for any fixed $j > 0$ and $m > 0$. Since $\{x_t\}$ is strictly stationary, ergodic and has finite and constant second moments, the almost sure convergence in (a) and (b) follows from the Ergodic Theorem (e.g. \citename{davidson94}, 1994, Theorem 13.12). Uniform convergence of the non-random term $\frac{1}{n} \sum_{i=1}^{n-1} \log f(\omega_i;\vtheta)$ follows by the same argument as in Zaffaroni (2009, Lemma 6). Finally, for all $\vtheta \in \Theta$ such that $\vtheta \neq \vtheta_0$, the first assumption in the proposition implies that $f(\omega;\vtheta_0)/f(\omega;\vtheta) - 1 > \log f(\omega;\vtheta_0) - \log f(\omega;\vtheta)$ on a set of positive Lebesgue measure, and hence $Q(\vtheta) > Q(\vtheta_0)$ unless $\vtheta = \vtheta_0$. \hfill $\blacksquare$

\subsection{Proof of Proposition 2}
Define $\delta_j = 1- \gamma_j$, $\eta_{j,i} = (m_{j,i} - \Exp(m)) - \delta_j (m_{j,i-1} - \Exp(m)$), $j=1,..,k$, $\eta_{0,i} = \log \epsilon_i - \Exp(\log \epsilon_1)$, $\vpsi_h = (\mathbf{1}_{\{h=0\}},\delta_1^h,\delta_2^h,...,\delta_k^h)$, $\veta_i = (\eta_{0,i}, \eta_{1,i},...,\eta_{k,i})'$, and $\mu = \bar{\psi} + \Exp(\log \epsilon_1) + k\Exp(m)$, and observe that $x_i$ can be written as
\begin{equation} \label{eq:marep}
x_i = \mu + \sum_{l=0}^{\infty} \vpsi_l \veta_{i-l}, \quad i \in \mathbb{Z}.
\end{equation}
Now define $\mathcal{F}_{j,i} = \sigma(\eta_{j,i},\eta_{j,i-1},...,)$, $j=0,...,k$ and observe that $\mathrm{E}|\eta_{j,i}| < \infty$, $\mathrm{E}(\eta_{j,i}|\mathcal{F}_{j,i-1}) = 0$ and $\mathrm{E}(\eta_{j,i}^2) = c < \infty$,
so $\{\eta_{j,i},\mathcal{F}_{j,i}\}$ is a homoskedastic martingale difference sequence, $j=1,...,k$. Now $\epsilon_i - \Exp(\epsilon_1)$ is \emph{iid} and the components of $\veta_i$ are independent, and hence $\{\veta_i,\mathcal{F}_i^{\eta}\}$, where $\mathcal{F}_{i}^{\eta} = \otimes_{j=0}^k \mathcal{F}_{j,i}$, is also homoskedastic martigale difference. Thus the statement in the proposition follows from \citename{ht82} (1982, Theorem 3.1) provided we verify that (a) the model spectral density satisfies the regularity conditions required by the theorem, and (b) the conditions (i)-(v) of the theorem are satisfied. Note that the non-zero constant term $\mu$ in (\ref{eq:marep}) is inconsequential since the periodogram is evaluated at Fourier frequencies, and hence the constant can be ignored in the sequel. If need be, it can be consistently estimated by the sample mean of $x_i$.

(a) \emph{Regularity conditions for spectral density.} We need to show that the model spectral density is bounded away from zero on $[-\pi,\pi]$, twice continuously differentiable function in $\vtheta$ on $\Theta$, and that the second derivatives are continuous in $\omega \in [-\pi,\pi]$. This follows directly from (\ref{eq:SDlogs}) and the fact that $\Theta$ is a compact subset of the parameter space (i.e. $\gamma_k$ is bounded away from zero and one and $b$ is bounded away from one).

(b.i) \emph{Conditional second moments.} We need to show that for each $j_1$, $j_2$ and $q$
\begin{equation} \label{eq:condition1}
\mathrm{Var}[\Exp(\eta_{j_1,i}\eta_{j_2,i+h}|\mathcal{F}_{i-q}^{\eta}) - 1_{\{j_1=j_2\}}\sigma_{\eta_{j_1}}^2] = O(q^{-2-r}), \quad r > 0,
\end{equation}
uniformly in $i$. Since the elements of $\eta_i$ are independent and second-order stationary, this follows immediately whenever $j_1 \neq j_2$.
For $j_1 = j_2$, we simplify notation by writing $j=j_1=j_2$ and further by writing $\overline{m}_{j,i} = m_{j,i} - \Exp(m)$, $\overline{m}_{j,i} = \overline{m}_i$, $\eta_{j,i} = \eta_i$ and $\delta = \delta_j$, $j=1,...,k$. The case of $j=0$ follows trivially since $\eta_{0,i}$ is \emph{iid}.
Now for all $h \geq 0$ and $p=1,...,4$,
\begin{equation} \label{eq:condmommp}
\Exp(\overline{m}_{i+h}^p|\overline{m}_i) = \delta^h \overline{m}_i^p + (1 - \delta^h)\mathrm{E}(\overline{m}^p), \\
\end{equation}
and for all $0 \leq h_1 < h_2$ and $q \geq 1$, by LIE,
\begin{eqnarray*}
\Exp(\overline{m}_{i+h_1} \overline{m}_{i+h_2}|\overline{m}_{i-q}) &=& \Exp(\overline{m}_{i+h_1}\Exp(\overline{m}_{i+h_2}|\overline{m}_{i+h_1})|\overline{m}_{i-q}), \\
                          &=& \delta^{h_2+q}\overline{m}_{i-q}^2 + \delta^{h_2-h_1}(1-\delta^{h_1+q})\Exp(m^2).
\end{eqnarray*}
For $h \geq 1$ and $q \geq 2$, we therefore have
\begin{eqnarray*}
\Exp(\eta_i \eta_{i+h}|\mathcal{F}_{i-q}^{\eta}) &=& \Exp(\overline{m}_i \overline{m}_{i+h}|\overline{m}_{i-q}) - \delta \Exp(\overline{m}_{i-1}\overline{m}_{i+h}|\overline{m}_{i-q}) \\
                                          && - \delta \Exp(\overline{m}_{i}\overline{m}_{i+h-1}|\overline{m}_{i-q}) + \delta^2 \Exp(\overline{m}_{i-1}\overline{m}_{i+h-1}|\overline{m}_{i-q}), \\
                                          &=& 0.
\end{eqnarray*}
For $h=0$ and any $q \geq 2$, we have
\begin{eqnarray*}
\Exp(\eta_i^2|\mathcal{F}_{i-q}^{\eta}) &=& \Exp(\overline{m}_i^2|\overline{m}_{i-q}) - 2\delta\Exp(\overline{m}_i \overline{m}_{i-1}|\overline{m}_{i-q}) + \delta^2 \Exp(\overline{m}_{i-1}^2|\overline{m}_{i-q}), \\
                                        &=& (\delta^q - \delta^{q+1})\overline{m}_{i-q}^2 + [(1-\delta^q) - \delta^2(1-\delta^{q-1})]\Exp(\overline{m}^2),
\end{eqnarray*}
and by fourth-order stationarity,
\begin{eqnarray*}
\mathrm{Var}[\Exp(\eta_i^2|\mathcal{F}_{i-q}^{\eta}) - \Exp(\eta_i^2)] &=& \mathrm{Var}[(\delta^q - \delta^{q+1})\overline{m}_{i-q}^2 + \{(1-\delta^q) - \delta^2(1-\delta^{q-1})\}\Exp(\overline{m}^2) -  (1-\delta^2)\Exp(\overline{m}^2)], \\
&=& (\delta^q - \delta^{q+1})^2 \mathrm{Var}(\overline{m}_{i-q}^2 - \Exp(\overline{m}^2)), \\
&=& O(\delta^{2q}),
\end{eqnarray*}
which verifies (\ref{eq:condition1}) since $O(\delta^{2q})$ decays faster than $O(q^{-2-r})$ for any $r < \infty$.

(b.ii) \emph{Conditional fourth-moments.} We need to show that for each $j_1,j_2,j_3,j_4$ and $h_1,h_2,h_3$,
\begin{equation} \label{eq:ii}
\Exp|\Exp(\eta_{j_1,i}\eta_{j_2,i+h_1}\eta_{j_3,i+h_2}\eta_{j_4,i+h_3}|\mathcal{F}_{i-q}) - \Exp(\eta_{j_1,i}\eta_{j_2,i+h_1}\eta_{j_3,i+h_2}\eta_{j_4,i+h_3})| = O(q^{-1-r}), \quad r > 0,
\end{equation}
uniformly in $i$, where $0\leq h_1 \leq h_2 \leq h_3$. To save space, we establish this for $j_1=j_2=j_3=j_4$, noting that the other cases can be shown analogously, and are easier since the elements of $\veta_i$ are independent. We again simplify notation by writing $j = j_1 = j_2 = j_3 = j_4$ , $\overline{m}_{j} = \overline{m}_{j,i}$ and $\delta = \delta_j$. Now
\begin{equation} \label{eq:auxeta4}
\eta_{i}\eta_{i+h_1}\eta_{i+h_2}\eta_{i+h_3} = (\overline{m}_i - \delta \overline{m}_{i-1})(\overline{m}_{i+h_1} - \delta \overline{m}_{i+h_1-1})(\overline{m}_{i+h_2} - \delta \overline{m}_{i+h_2-1})(\overline{m}_{i+h_3} - \delta \overline{m}_{i+h_3-1}),
\end{equation}
where for any $0\leq h_1 \leq h_2 \leq h_3$ the sum on the right-hand side contains sixteen terms, each being a product of four terms with either zero, one, two, three or four lags in common depending on the particular choice of $h_1,h_2$ and $h_3$. Starting with the case when all the lags are distinct, we obtain by iterated conditioning and repeated use of (\ref{eq:condmommp})
\begin{eqnarray*}
\Exp(\overline{m}_{i}\overline{m}_{i+l_1}\overline{m}_{i+l_2}\overline{m}_{i+l_3}|\overline{m}_{i-q}) &=& \delta^{l_3}[\delta^q \overline{m}_{i-q}^4 + (1-\delta^q)\Exp(\overline{m}^4)] + \delta^{l_3-l_1}(1-\delta^{l_1})\Exp(\overline{m}^3)\delta^q \overline{m}_{i-q} \\ && + \delta^{l_3-l_2+l_1}(1-\delta^{l_2-l_1})\Exp(\overline{m}^2)[\delta^q \overline{m}_{i-q}^2 + (1-\delta^q)\Exp(\overline{m}^2)] \\
\Exp(\overline{m}_{i}\overline{m}_{i+l_1}\overline{m}_{i+l_2}\overline{m}_{i+l_3}) &=& \delta^{l_3} \Exp(\overline{m}^4) + \delta^{l_3-l_2+l_1}(1-\delta^{l_2-l_1})\Exp(\overline{m}^2)^2, \\
&& \hspace{6.5cm} 0 < l_1 < l_2 < l_3, q \geq 2.
\end{eqnarray*}
Thus by fourth-order stationarity
\begin{eqnarray*}
&& \hspace{-2cm}\Exp|\Exp(\overline{m}_{i}\overline{m}_{i+l_1}\overline{m}_{i+l_2}\overline{m}_{i+l_3}|\overline{m}_{i-q}) - \Exp(\overline{m}_{i}\overline{m}_{i+l_1}\overline{m}_{i+l_2}\overline{m}_{i+l_3})| \\
&& \hspace{2cm} \leq \delta^q \{\delta^{l_3}\Exp|\overline{m}_{i-q}^4 - \Exp(\overline{m}^4)| + \delta^{l_3-l_1}(1-\delta^{l_1})|\Exp(\overline{m}^3)|\Exp|\overline{m}_{i-q}|\} \\
&& \hspace{2.5cm}  + \delta^{l_3-l_2+l_1}(1-\delta^{l_2-l_1})\Exp(m^2)\Exp|\overline{m}_{i-q}^2 - \Exp(\overline{m}^2)|  \\
&& \hspace{2cm} =  O(\delta^q).
\end{eqnarray*}
Turning to the case when all the lags are equal, i.e. $0 = l_1 = l_2 = l_3$, we have
\begin{eqnarray*}
\Exp|\Exp(\overline{m}m_{i}^4|\overline{m}_{i-q}) - \Exp(\overline{m}_{i}^4)| &\leq& \delta^q \Exp|\overline{m}_{i-q}^4 - \Exp(\overline{m}^4)|, \\
&=& O(\delta^q).
\end{eqnarray*}
The terms with one, two and three common lags in (\ref{eq:auxeta4}) can be established analogously. (\ref{eq:ii}) then follows from the triangle inequality and the fact that $O(\delta^q)$ decays faster than $O(q^{-1-r})$ for any $r < \infty$.

(b.iii) \emph{Square-integrability of model spectral density.} We need to show that the MSMD spectral density is square-integrable. This follows directly from the continuity and boundedness of the spectral density on $[-\pi,\pi]$, uniformly on $\Theta$.

(b.iv) \emph{Absolute summability of fourth-order cumulants.} We need to show that
\begin{equation} \label{eq:sumcum}
\sum_{h_1=-\infty}^{\infty}\sum_{h_2=-\infty}^{\infty}\sum_{h_3=-\infty}^{\infty} |\mathrm{cum}(\eta_{j_1,i},\eta_{j_2,i+h_1},\eta_{j_3,i+h_2},\eta_{j_4,i+h_3})| < \infty
\end{equation}
for any $j_1,j_2,j_3,j_4$, $j_l = 0,...,k$ and $l=1,2,3,4$. Since the elements of $\veta_i$ are independent and fourth-order stationary all cross-cumulants are zero, so we only need to focus on the case $j_1=j_2=j_3=j_4$. Now since $\eta_{0,i}$ is \emph{iid} with finite fourth moment, the absolute summability of its fourth-order cumulants follows directly. For an $\eta_{j,i}$, $j=1,...,k$, we use the fact that $\eta_{j,i}$ is a fourth-order stationary exponential strong mixing with $4 + r$ moment finite and hence $\sum_{s=0}^{\infty} (s+1)^{p-2}\alpha(s)^{r/(p+r)} < \infty$, $p=1,...,4$, where $\alpha(s)$ is the strong mixing coefficient associated to $\eta_{j,i}$ satisfying $\alpha(s) = O(\rho^s)$ for some $\rho \in (0,1)$. Thus by \citename{dl89} (1989, Proposition 2.2), the inequality (\ref{eq:sumcum}) holds.

(b.v) \emph{Lipschitz continuity of $f(\omega;\vtheta_0)$.} This follows directly for the smoothness of the first derivatives of $f(\omega;\vtheta)$ uniformly on $\Theta$.

The statement of the proposition now follows by simplifying the formulae for $\mM_f$ and $\tilde{\mV}$ in \citename{ht82} (1982, Theorem 3.1) in view of the fact that the true spectral density of $x_i$ is given by $f(\omega;\vtheta_0)$, together with the fact that the matrix $\mM(\vtheta_0)$ is full rank by the identification assumption stated in Proposition 1. \hfill $\blacksquare$

\subsection{Proof of Proposition 3}
We need to show that the assumptions of Theorem 1 in \citeasnoun{dhsw09} are satisfied. Assumption (ii) requires the duration process to be exponential strong mixing, and this was established in Section 2.1; it is well-known that exponential $\beta$-mixing implies exponential strong mixing since the $\beta$-mixing coefficients dominate the strong mixing coefficients, e.g. \citename{davidson94}, 1994, Chapter 14.

Turning to Assumption (i) of Theorem 1 in \citeasnoun{dhsw09}, we first note that the long-run variance of $X_i$ is strictly positive, i.e. $\lim_{n \rightarrow \infty} \frac{1}{n} \Var(\sum_{i=1}^n X_i) = \sigma^2 > 0$ for some constant $\sigma^2$. Since $\{X_i\}$ is stationary with exponentially declining autocovariances, this is equivalent to showing that the spectral density of $X_i$ is strictly positive at the zero frequency. This follows directly from (\ref{eq:sdXl}) since when $\omega = 0$ each summand in (\ref{eq:sdXl}) is strictly positive implying $2\pi f_X(\omega) > 0$. Write $\sigma^2 = 2\pi f_X(\omega)$ and define $Y_n(s) = n^{-1/2} \sum_{i=1}^{\lfloor ns \rfloor} (X_i - \bar{\psi})$. Since by assumption $\epsilon_i$ is \emph{iid} with $3+r$ moments finite for some $r > 0$ and independent of $\psi_i$, which also has $3+r$ moments finite, it follows that $X_i$ has $3+r$ moments finite. Then by \citeasnoun{dhsw09}, $Y_n \Rightarrow \sigma W$, as $n \rightarrow \infty$, where $W$ is a standard Brownian motion. This verifies Assumption (i) of Theorem 1 in \citeasnoun{dhsw09}.

Finally, by the same argument as in the proof of Theorem 3 in \citeasnoun{dhsw09}, the exponential strong mixing property and the existence of $3+r$ (where $r > 0$) moments of $x_i$ imply for $y_n = n^{-1/2} \sum_{i=1}^n x_i$, and some constant $c$ that $\Exp(|y_n - \Exp(y_n)|^{3+r}) \leq c < \infty$ as required by Assumption (iii) of Theorem 1 in \citeasnoun{dhsw09}. \hfill $\blacksquare$

\clearpage
\section{Tables and Figures}

\begin{table}[!h]
\begin{center}
\begin{tabular}{lrrrrrrrrr}
\toprule
    &         \multicolumn{4}{c}{Exp}      & & \multicolumn{4}{c}{$\mathcal{W}(\kappa)$} \\
    \cmidrule{2-5} \cmidrule{7-10}
    & 1,000 & 2,500 & 5,000 & 10,000 & &   1,000 & 2,500 & 5,000 & 10,000 \\
\midrule
\multicolumn{5}{l}{A. Binomial multipliers - MLE}  \\
$m_0$         &  1.382  & 1.393  & 1.395 &  -  & &  1.393 &  1.3993 &  1.400 &   -  \\
              & (0.035)  & (0.021)  & (0.016) &  ( - )  & &  (0.036) &  (0.022) &  (0.015) &  ( - ) \\
$b$           & 1.832  & 1.936  & 1.949 &  -  & &  1.982 &  1.998 &  2.022 &   -  \\
              & (0.346)  & (0.250)  & (0.162) &  ( - )  & &  (0.438) &  (0.258) &  (0.180) &  ( - ) \\
$\gamma_k$    & 0.501  & 0.499  & 0.494 &  -  & &  0.506 &  0.502 &  0.509 &   -  \\
              & (0.164)  & (0.107)  & (0.069) &  ( - )  & &  (0.160) &  (0.101) &  (0.064) &  ( - ) \\
$\kappa$      &     -   &    -    &    -  &  -   & &  1.465 &  1.458 &  1.453 &   -  \\
              &     ( - )   &    ( - )    &   ( - )  &  ( - ) & &  (0.098) &  (0.064) &  (0.037) &  ( - ) \\
\midrule
\multicolumn{5}{l}{B. Binomial multipliers - Whittle}  \\
$m_0$      & 1.425 & 1.409 & 1.400 & 1.400 & & 1.437 & 1.411 & 1.403 & 1.401 \\
           & (0.080) & (0.047) & (0.018) & (0.007) & & (0.088) & (0.045) & (0.019) & (0.013) \\
$b$        & 2.608 & 2.231 & 1.996 & 1.999 & & 2.883 & 2.222 & 2.022 & 2.012 \\
           & (1.871) & (1.190) & (0.214) & (0.131) & & (2.263) & (1.145) & (0.247) & (0.152) \\
$\gamma_k$ & 0.467 & 0.497 & 0.499 & 0.502 & & 0.514 & 0.524 & 0.523 & 0.514 \\
           & (0.221) & (0.146) & (0.101) & (0.075) & & (0.273) & (0.208) & (0.158) & (0.104) \\
$\kappa$   & -  &  -  &  -  &  -  & & 1.617 & 1.523 & 1.491 & 1.466 \\
           & ( - ) & ( - ) & ( - ) & ( - ) & & (0.415) & (0.244) & (0.161) & (0.098) \\
\midrule
\multicolumn{5}{l}{C. Log-normal multipliers - Whittle}  \\
$\lambda$  & 0.206  & 0.167 & 0.153 & 0.150 & & 0.214 & 0.168 & 0.155 & 0.151 \\
           & (0.143)& (0.076) & (0.031) & (0.015) & & (0.137) & (0.068) & (0.030) & (0.015) \\
$b$        & 2.803  & 2.279 & 2.034 & 1.994 & & 2.924 & 2.257 & 2.046 & 2.008 \\
           &(2.089) & (1.282) & (0.483) & (0.182) & & (2.267) & (1.216) & (0.455) & (0.195) \\
$\gamma_k$ & 0.474 & 0.498 & 0.501 & 0.499 & & 0.525 & 0.531 & 0.528 & 0.516 \\
           & (0.247) & (0.173) & (0.124) & (0.086) & & (0.296) & (0.225) & (0.176) & (0.128) \\
$\kappa$   &  -  &  -  &  -  &  -  & & 1.583 & 1.515 & 1.487 & 1.465 \\
           & ( - ) & ( - ) & ( - ) & ( - ) & & (0.315) & (0.192) & (0.137) & (0.075) \\
\bottomrule
\end{tabular}
\caption{Monte Carlo simulation of the maximum likelihood (MLE) and Whittle estimators of the parameters of the MSMD model with $k=8$ binomial or log-normal multipliers and exponential or Weibull innovations. We report average parameter estimates obtained in the simulation together with standard errors in parentheses. The true parameters used in the simulations are $b=2$, $\gamma_k=0.5$, $m_0=1.4$, $\lambda =0.15$ and $\kappa=1.45$. The results for MLE are based on 500, 200 and 100 replications for the samples of $n=1,000$, $2,000$ and $5,000$ observations, respectively. All simulations of the Whittle estimator are based on 1,000 replications. \label{tab:mcsim}}
\end{center}
\end{table}

\clearpage
\begin{table}
\begin{center}
\begin{tabular}{lrrr}
\toprule
  &  \multicolumn{3}{c}{Distribution of $\epsilon$} \\
  \cmidrule{2-4}
  &  exp & $\mathcal{W}(\kappa)$ & LN($\lambda$) \\
\midrule
\multicolumn{4}{l}{A. 10\% nominal} \\
Binomial multipliers  & 11.3  & 9.2  & 9.9  \\
Lognormal multipliers & 11.0  & 11.7  & 13.6  \\
\midrule
\multicolumn{4}{l}{B. 5\% nominal} \\
Binomial multipliers  & 6.2  & 6.0  & 5.6  \\
Lognormal multipliers & 6.2  & 6.4  & 7.0  \\
\bottomrule
\end{tabular}
\caption{Monte Carlo simulation of the goodness-of-fit test for the MSMD model with $k=8$ binomial or log-normal multipliers and exponential, Weibull or log-normal innovations. We report the simulated size for 5 and 10\% nominal level. The true parameters used in the simulations are $b=2$, $\gamma_k=0.5$, $m_0=1.4$, $\lambda =0.15$ and $\kappa=1.45$. The sample size used in the simulations is $n=10,000$ observations and the simulations are based on 1,000 replications. \label{tab:mcsimgf}}
\end{center}
\end{table}

\begin{table}
\small
\centering				
\begin{tabular}{lrrrrrrrrrrr}	
\toprule
 & \multicolumn{3}{c}{CHF} & & \multicolumn{3}{c}{EUR} & & \multicolumn{3}{c}{JPY}	  \\
\cmidrule{2-4}\cmidrule{6-8}\cmidrule{10-12}  									
 & raw & adj & log-adj & & raw & adj & log-adj & & raw & adj & log-adj  \\
\midrule 									
Mean & 117.8 & 1.000 & -0.602 & & 105.5 & 1.002 & -0.580  & & 89.50 & 1.002 &  -0.720\\
Median & 66.00 & 0.618 &-0.481  & & 59.00 & 0.619 & -0.479  & & 43.00 & 0.536 & -0.623 \\
Minimum & 1.000 & 0.004 & -5.613 & & 1.000 & 0.004 & -5.482 & & 1.000  & 0.007 & -5.028 \\
Maximum & 2498 & 15.78 & 2.759  & & 3574 & 20.55 & 3.023 & & 3160 & 20.31 & 3.011 \\
Std.dev. &  159.1	& 1.194	& 1.204	& & 141.1	& 1.205	& 1.167	& & 135.3	& 1.386	& 1.296 \\
Dispersion &  1.351 & 1.194 & - & & 1.337 & 1.204 & - & & 1.512 & 1.383 & - \\
Skewness &   3.775	& 3.307	& -0.497	& & 4.410	& 3.731	& -0.439	& & 4.586	& 4.053	& -0.322 \\
Kurtosis &  26.43	& 22.10	& 3.187	& & 49.99	& 29.71	& 3.140	& & 46.06	& 30.60	& 2.817 \\
Left tail &    -    &  -    & 10.16  & &   -   &   -  & 9.409   & &  -    &  -  &  12.88   \\
           &   -   &   -    & (0.002)  & &  -   &   - & (0.002)   & &  -  &  - & (0.002)   \\
Right tail &   -     &  3.976    &  9.703 & &  -    &  3.501   &  7.949  & &   -   &  3.206  & 7.955   \\
           &   -    &  (0.003)   & (0.002)  & &  -    &  (0.003) & (0.003)   & &  -  &  (0.003) & (0.003)   \\
\bottomrule	
\end{tabular}	
\caption{Descriptive statistics for Swiss franc (CHF), Euro (EUR) and Japanese Yen (JPY) futures price durations. The sample period runs between 9 November 2009 and 29 January 2010 and the sample size is $n=12,000$ for all datasets. The columns labeled ``raw" report descriptive statistics for the raw price durations, the columns ``adj" give descriptive statistics for the seasonally adjusted durations and the columns ``log-adj" report descriptive statistics for the logarithmic seasonally adjusted durations. The rows labeled ``Left tail" and ``Right tail" report the estimated tail indexes (with standard error in parentheses) for the left and right tail, respectively. \label{tab:descriptivestats}}	
\end{table}		

\clearpage
	
\begin{table}
\small
\centering				
\begin{tabular}{lrrrrrrr}	
\toprule
 & \multicolumn{3}{c}{Exp} & &  \multicolumn{3}{c}{$\mathcal{W(\kappa)}$} \\
 \cmidrule{2-4}  \cmidrule{6-8}
              & $k=4$ & $k=6$ & $k=8$ & & $k=4$ & $k=6$ & $k=8$ \\
 \midrule
\multicolumn{8}{l}{A. Binomial multipliers - MLE} \\			
$m_0$ &1.356 & 1.328 & 1.316 & & 1.439 & 1.394 & 1.355 \\
 & (0.009) & (0.008) & (0.010) & & (0.008) & (0.009) & (0.007) \\
$b$ &2.339 & 1.962 & 2.045 & & 7.059 & 6.075 & 4.188 \\
 & (0.551) & (0.311) & (0.308) & &  (0.779) & (0.621) & (0.336) \\
$\gamma_k$   &0.067 & 0.059 & 0.064 & & 0.533 & 0.973 & 0.999 \\
      & (0.014) & (0.013) & (0.014) & & (0.062) & (0.031) & (0.002) \\
$\kappa$ & -  &  -  &  -  & & 1.346 & 1.485 & 1.566 \\
           & ( - ) & ( - ) & ( - ) & & (0.024) & (0.041) & (0.045) \\
$T_n$ &20.078 & 22.644 & 24.322 & & -1.150 & -1.252 & -1.295 \\
 &(0.000) & (0.000) & (0.000) & & (0.875) & (0.895) & (0.902) \\
$\log L$   &-8881.967 & -8859.848 & -8863.670 & & -8661.562 & -8614.860 & -8594.576 \\
\midrule
\multicolumn{8}{l}{B. Binomial multipliers - Whittle} \\
$m_0$ &1.293 & 1.243 & 1.212 & & 1.369 & 1.308 & 1.269 \\
 & (0.014) & (0.012) & (0.010) & & (0.017) & (0.017) & (0.015) \\
$b$ &1.876 & 1.477 & 1.326 & & 4.150 & 2.600 & 2.045 \\
 & (0.222) & (0.121) & (0.083) & & (0.836) & (0.449) & (0.250) \\
$\gamma_k$   &0.058 & 0.058 & 0.058 & & 0.476 & 0.568 & 0.608 \\
      & (0.015) & (0.016) & (0.016) & & (0.213) & (0.299) & (0.287) \\
$\kappa$ & -  &  -  &  -  & & 1.376 & 1.380 & 1.379 \\
           & ( - ) & ( - ) & ( - ) & & (0.047) & (0.059) & (0.054) \\

$T_n$ &35.836 & 36.270 & 36.480 & & -1.084 & -1.079 & -1.077 \\
 &(0.000) & (0.000) & (0.000) & & (0.861) & (0.860) & (0.859) \\
q-log $L$   &-2923.704 & -2924.308 & -2924.557 & & -3381.884 & -3381.712 & -3381.651 \\
\midrule
\multicolumn{8}{l}{C. Log-normal multipliers - Whittle} \\
$\lambda$ &0.096 & 0.064 & 0.048 & & 0.162 & 0.107 & 0.079 \\
 & (0.010) & (0.007) & (0.005) & & (0.017) & (0.014) & (0.009) \\
$b$ &1.876 & 1.477 & 1.326 & & 4.150 & 2.600 & 2.045 \\
 & (0.222) & (0.121) & (0.083) & & (0.836) & (0.449) & (0.250) \\
$\gamma_k$   &0.058 & 0.058 & 0.058 & & 0.476 & 0.568 & 0.608 \\
      & (0.015) & (0.016) & (0.016) & & (0.213) & (0.299) & (0.287) \\
$\kappa$ & -  &  -  &  -  & & 1.376 & 1.380 & 1.379 \\
           & ( - ) & ( - ) & ( - ) & & (0.047) & (0.059) & (0.054) \\
$T_n$ &35.835 & 36.270 & 36.479 & & -1.084 & -1.078 & -1.077 \\
 & (0.000) & (0.000) & (0.000) & & (0.861) & (0.860) & (0.859) \\
q-log $L$  &-2923.704 & -2924.308 & -2924.557 & & -3381.884 & -3381.712 & -3381.651 \\
\bottomrule	
\end{tabular}	
\caption{MSMD parameter estimates for de-seasonalised CHF price durations. (A) Maximum likelihood estimates (MLE) of the binomial MSMD models with exponential and Weibull innovations, (B) Whittle estimates of the binomial MSMD models with exponential and Weibull innovations and (C) Whittle estimates of the log-normal MSMD models with exponential and Weibull innovations. Standard errors are reported in parentheses. The specification test $T_n$ is reported with p-values in parentheses. \label{tab:MSMDparametersCHF}}	
\end{table}	

\clearpage
	
\begin{table}
\small	
\centering				
\begin{tabular}{lrrrrrrr}	
\toprule
 & \multicolumn{3}{c}{Exp} & &  \multicolumn{3}{c}{$\mathcal{W(\kappa)}$} \\
 \cmidrule{2-4}  \cmidrule{6-8}
           & $k=4$ & $k=6$ & $k=8$ & & $k=4$ & $k=6$ & $k=8$ \\
 \midrule
\multicolumn{8}{l}{A. Binomial multipliers - MLE} \\			
$m_0$ &1.349 & 1.281 & 1.331 & & 1.452 & 1.397 & 1.342 \\
 & (0.011) & (0.011) & (0.020) & & (0.007) & (0.008) & (0.008) \\
$b$ &3.001 & 1.932 & 3.000 & & 7.241 & 5.491 & 3.289 \\
 & (0.636) & (0.299) & (0.525) & &  (0.737) & (1.032) & (0.563) \\
$\gamma_k$   &0.086 & 0.088 & 0.089 & & 0.881 & 0.999 & 0.999 \\
      & (0.015) & (0.022) & (0.017) & & (0.032) & (0.004) & (0.004) \\
$\kappa$ & -  &  -  &  -  & & 1.554 & 1.675 & 1.670 \\
           & ( - ) & ( - ) & ( - ) & & (0.029) & (0.068) & (0.067) \\
$T_n$ &33.655 & 36.987 & 36.923 & & -0.935 & -0.975 & -1.041 \\
 &(0.000) & (0.000) & (0.000) & & (0.825) & (0.835) & (0.851) \\
$\log L$   &-9007.010 & -8998.909 & -9001.426 & & -8706.568 & -8652.807 & -8649.950 \\
\midrule
\multicolumn{8}{l}{B. Binomial multipliers - Whittle} \\
$m_0$ &1.272 & 1.224 & 1.195 & & 1.402 & 1.354 & 1.308 \\
 & (0.013) & (0.011) & (0.010) & & (0.062) & (0.023) & (0.017) \\
$b$ &1.643 & 1.353 & 1.242 & & 5.060 & 3.391 & 2.459 \\
 & (0.343) & (0.197) & (0.139) & & (2.837) & (0.585) & (0.260) \\
$\gamma_k$   &0.058 & 0.058 & 0.058 & & 0.901 & 0.999 & 0.999 \\
      & (0.017) & (0.019) & (0.021) & & (0.438) & (0.008) & (0.005) \\
$\kappa$ & -  &  -  &  -  & & 1.588 & 1.719 & 1.701 \\
           & ( - ) & ( - ) & ( - ) & & (0.302) & (0.173) & (0.136) \\
$T_n$ &53.333 & 53.720 & 53.887 & & -1.614 & -1.619 & -1.619 \\
 &(0.000) & (0.000) & (0.000) & & (0.947) & (0.947) & (0.947) \\
q-log $L$   &-3071.748 & -3071.930 & -3072.001 & & -3648.922 & -3648.827 & -3648.782 \\
\midrule
\multicolumn{8}{l}{C. Log-normal multipliers - Whittle} \\
$\lambda$ &0.081 & 0.053 & 0.040 & & 0.199 & 0.147 & 0.106 \\
 & (0.008) & (0.005) & (0.004) & & (0.076) & (0.023) & (0.014) \\
$b$ &1.643 & 1.352 & 1.242 & & 5.060 & 3.391 & 2.459 \\
 & (0.343) & (0.197) & (0.139) & & (2.835) & (0.585) & (0.263) \\
$\gamma_k$   &0.058 & 0.058 & 0.058 & & 0.901 & 0.999 & 0.999 \\
      & (0.017) & (0.019) & (0.021) & & (0.438) & (0.008) & (0.005) \\
$\kappa$ & -  &  -  &  -  & & 1.588 & 1.719 & 1.701 \\
           & ( - ) & ( - ) & ( - ) & & (0.301) & (0.173) & (0.139) \\
$T_n$ &53.333 & 53.720 & 53.888 & & -1.614 & -1.619 & -1.619 \\
 & (0.000) & (0.000) & (0.000) & & (0.947) & (0.947) & (0.947) \\
q-log $L$  &-3071.748 & -3071.930 & -3072.001 & & -3648.922 & -3648.827 & -3648.782 \\
\bottomrule	
\end{tabular}	
\caption{MSMD parameter estimates for de-seasonalised Euro futures price durations. (A) Maximum likelihood estimates (MLE) of the binomial MSMD models with exponential and Weibull innovations, (B) Whittle estimates of the binomial MSMD models with exponential and Weibull innovations and (C) Whittle estimates of the log-normal MSMD models with exponential and Weibull innovations. Standard errors are reported in parentheses. The specification test $T_n$ is reported with p-values in parentheses. \label{tab:MSMDparametersEUR}}	
\end{table}	

\clearpage
	
\begin{table}
\small	
\centering				
\begin{tabular}{lrrrrrrr}	
\toprule
 & \multicolumn{3}{c}{Exp} & &  \multicolumn{3}{c}{$\mathcal{W(\kappa)}$} \\
 \cmidrule{2-4}  \cmidrule{6-8}
            & $k=4$ & $k=6$ & $k=8$ & & $k=4$ & $k=6$ & $k=8$ \\
 \midrule
\multicolumn{8}{l}{A. Binomial multipliers - MLE} \\			
$m_0$ &1.415 & 1.343 & 1.343 & & 1.481 & 1.424 & 1.376 \\
 & (0.010) & (0.014) & (0.013) & & (0.007) & (0.007) & (0.008) \\
$b$ &3.348 & 2.188 & 2.454 & & 4.848 & 4.006 & 2.933 \\
 & (0.552) & (0.285) & (0.263) & &  (0.366) & (0.340) & (0.253) \\
$\gamma_k$   &0.152 & 0.164 & 0.180 & & 0.826 & 0.985 & 0.999 \\
      & (0.018) & (0.025) & (0.025) & & (0.047) & (0.012) & (0.002) \\
$\kappa$ & -  &  -  &  -  & & 1.481 & 1.590 & 1.672 \\
           & ( - ) & ( - ) & ( - ) & & (0.035) & (0.040) & (0.063) \\
$T_n$ &26.401 & 28.896 & 30.695 & & 1.006 & 1.508 & 1.459 \\
 &(0.000) & (0.000) & (0.000) & & (0.157) & (0.066) & (0.072) \\
$\log L$   &-8255.173 & -8252.226 & -8256.733 & & -8053.231 & -8008.807 & -8000.278 \\
\midrule
\multicolumn{8}{l}{B. Binomial multipliers - Whittle} \\
$m_0$ &1.342 & 1.283 & 1.247 & & 1.434 & 1.367 & 1.323 \\
 & (0.013) & (0.011) & (0.010) & & (0.012) & (0.015) & (0.015) \\
$b$ &1.991 & 1.560 & 1.391 & & 3.954 & 2.574 & 2.046 \\
 & (0.315) & (0.157) & (0.103) & & (0.422) & (0.285) & (0.185) \\
$\gamma_k$   &0.084 & 0.091 & 0.095 & & 0.569 & 0.723 & 0.784 \\
      & (0.014) & (0.016) & (0.018) & & (0.079) & (0.170) & (0.183) \\
$\kappa$ & -  &  -  &  -  & & 1.396 & 1.422 & 1.430 \\
           & ( - ) & ( - ) & ( - ) & & (0.031) & (0.063) & (0.073) \\
$T_n$ &49.062 & 49.129 & 49.138 & & 0.757 & 0.748 & 0.745 \\
 &(0.000) & (0.000) & (0.000) & & (0.225) & (0.227) & (0.228) \\
q-log $L$ &-2501.523 & -2501.468 & -2501.450 & & -2883.415 & -2883.466 & -2883.507 \\
\midrule
\multicolumn{8}{l}{C. Log-normal multipliers - Whittle} \\
$\lambda$ &0.136 & 0.088 & 0.065 & & 0.241 & 0.160 & 0.119 \\
 & (0.012) & (0.008) & (0.007) & & (0.017) & (0.016) & (0.012) \\
$b$ &1.991 & 1.560 & 1.391 & & 3.954 & 2.574 & 2.046 \\
 & (0.315) & (0.157) & (0.125) & & (0.422) & (0.284) & (0.185) \\
$\gamma_k$   &0.084 & 0.091 & 0.095 & & 0.569 & 0.723 & 0.784 \\
      & (0.014) & (0.016) & (0.011) & & (0.079) & (0.170) & (0.183) \\
$\kappa$ & -  &  -  &  -  & & 1.396 & 1.422 & 1.430 \\
           & ( - ) & ( - ) & ( - ) & & (0.031) & (0.063) & (0.073) \\
$T_n$ &49.061 & 49.128 & 49.514 & & 0.757 & 0.748 & 0.745 \\
 & (0.000) & (0.000) & (0.000) & & (0.225) & (0.227) & (0.228) \\
q-log $L$ &-2501.523 & -2501.468 & -2501.347 & & -2883.415 & -2883.466 & -2883.507 \\
\bottomrule	
\end{tabular}		
\caption{MSMD parameter estimates for de-seasonalised Japanese Yen futures price durations. (A) Maximum likelihood estimates (MLE) of the binomial MSMD models with exponential and Weibull innovations, (B) Whittle estimates of the binomial MSMD models with exponential and Weibull innovations and (C) Whittle estimates of the log-normal MSMD models with exponential and Weibull innovations. Standard errors are reported in parentheses. The specification test $T_n$ is reported with p-values in parentheses. \label{tab:MSMDparametersJPY}}
\end{table}	

\clearpage
	
\begin{table}
\small	
\centering				
\begin{tabular}{lrrrrr}	
\toprule									
 & \multicolumn{2}{c}{ACD} & &   \multicolumn{2}{c}{LMSD} \\
\cmidrule{2-3}\cmidrule{5-6}   									
 & Exp & $\mathcal{W(\kappa)}$ & & Exp &$\mathcal{W(\kappa)}$   \\
\midrule
\multicolumn{6}{l}{A. Swiss Franc}	\\								
$\alpha$      &0.152 & 0.151 & &  -  &  -  \\
              &(0.005) & (0.000) & &( - ) & ( - ) \\
$\beta$       &0.830 & 0.858 & & 0.879 & -0.043\\
              &(0.007) & (0.000) & & (0.043) & (0.099) \\
$d$           & -  &  -  & & 0.500 & 0.477 \\
              &( - ) & ( - ) & & (0.074) & (0.033) \\
$\sigma_u^2$  & -  &  -  & & 0.005 & 0.382 \\
              &( - ) & ( - ) & & (0.001) & (0.143)\\
$\kappa$      & -  & 0.996 & &  -  & 1.519 \\
              &( - ) & (0.002) & & ( - ) & (0.125) \\
$\log L$      &-8953.300 & -8930.917 &q-log $L$ & -2928.500 & -3376.700 \\
\midrule
\multicolumn{6}{l}{B. Euro} \\
$\alpha$      &0.157 & 0.155 & & -  &  - \\
              &(0.006) & (0.006) & &( - ) & ( - ) \\
$\beta$       & 0.811 & 0.812 & &0.850 & -0.054 \\
              &(0.008) & (0.007) & & (0.054) & (0.046) \\
$d$           &  -  &  -  & &0.500 & 0.387 \\
              & ( - ) & ( - ) & & (0.073) & (0.037) \\
$\sigma_u^2$  &  -  &  -  & &0.006 & 0.640 \\
              & ( - ) & ( - ) & & (0.002) & (0.186) \\
$\kappa$      &   -  & 1.034 & & -  & 1.951 \\
              & ( - ) & (0.007) & & ( - ) & (0.369) \\
$\log L$      &-9068.800 & -9058.900 &q-log $L$ &-3074.600 & -3643.700 \\
\midrule
\multicolumn{6}{l}{C. Japanese Yen} \\
$\alpha$      &0.188 & 0.193 & & -  &  -  \\
              &(0.005) & (0.006) & &( - ) & ( - ) \\
$\beta$       & 0.780 & 0.775 & &0.848 & -0.081 \\
              &(0.007) & (0.008) & & (0.043) & (0.025) \\
$d$           &   -  &  -  & &0.367 & 0.373 \\
              & ( - ) & ( - ) & & (0.076) & (0.034) \\
$\sigma_u^2$  &  -  &  -  & & 0.020 & 1.022 \\
              & ( - ) & ( - ) & & (0.004) & (0.231)  \\
$\kappa$      & -  & 0.949 & & -  & 2.570 \\
              & ( - ) & (0.007) & & ( - ) & (1.067) \\
$\log L$      & -8500.100 & -8473.900 &q-log $L$ &-2515.400 & -2888.600 \\	
\bottomrule
\end{tabular}		
\caption{Maximum likelihood estimates of exponential and Weibull ACD models and Whittle estimates of exponential and Weibull LMSD models for de-seasonalised (A) Swiss franc, (B) Euro and (C) Japanese Yen futures price durations. Standard errors are reported in parentheses. \label{tab:ACDLMSDparameters}}
\end{table}		

\clearpage
\begin{landscape}
\begin{table}
\footnotesize
\centering				
\begin{tabular}{llllllllllll}	
\toprule							
 & \multicolumn{2}{c}{$h=1$} & &   \multicolumn{2}{c}{$h=5$} & &  \multicolumn{2}{c}{$h=10$} & &  \multicolumn{2}{c}{$h=20$}	  \\
\cmidrule{2-3}\cmidrule{5-6}\cmidrule{8-9}\cmidrule{11-12}				
 & Exp & $\mathcal{W(\kappa)}$ & & Exp & $\mathcal{W(\kappa)}$ & & Exp & $\mathcal{W(\kappa)}$ & & Exp & $\mathcal{W(\kappa)}$ \\
\midrule	
\multicolumn{8}{l}{A. Swiss franc} \\
ACD & 1.289 & 1.243 & & 11.652 & 9.815 & &
39.372 & 28.899 & & 159.352 & 91.237 \\LMSD &
1.213** & 1.248  & & 8.900** & 9.061 & & 25.760** & 25.806* & & 79.218** & 77.916* \\MSMD(6) MLE&
1.221** & 1.206* & & 8.843** & 8.378** & & 23.981** & 23.992* & & 77.708** & 78.975 \\MSMD(8) MLE&
1.219** & 1.215 & & 8.781** & 8.292** & & 23.628** & 23.109** & & 75.082** & 76.671* \\MSMD(6) Bin-Whittle&
1.235** & 1.269 & & 9.645** & 10.088  & & 27.539** & 27.989  & & 85.064** & 84.511 \\MSMD(8) Bin-Whittle&
1.236** & 1.270 & & 9.656** & 10.102  & & 27.583** & 28.016  & & 85.250** & 84.565 \\MSMD(6) Log-Whittle&
1.236** & 1.273 & & 9.682** & 10.147  & & 27.658** & 28.097  & & 85.467** & 84.721 \\MSMD(8) Log-Whittle&
1.236** & 1.273 & & 9.684** & 10.147  & & 27.674** & 28.098  & & 85.560** & 84.720 \\MSMD(6)+MSMD(8)+LMSD MLE&
1.211** & 1.213** & & 8.708** & 8.427** & & 23.939** & 23.202** & & 74.231** & 73.393** \\MSMD(6)+MSMD(8)+LMSD Bin-Whitt.&
1.215** & 1.224** & & 8.672** & 8.503** & & 26.015** & 25.779** & & 80.372** & 78.770** \\MSMD(6)+MSMD(8)+LMSD Log-Whitt.&
1.216** & 1.225** & & 8.691** & 8.346** & & 26.072** & 25.820** & & 80.561** & 78.852** \\
\midrule	
\multicolumn{8}{l}{B. Euro} \\
ACD & 1.636 & 1.635 & & 13.495 & 13.476 & &
37.628 & 37.565 & & 106.023 & 105.897 \\LMSD &
1.570** & 1.577* & & 12.219** & 12.424* & & 32.266** & 32.501** & & 86.183** & 86.461** \\MSMD(6) MLE&
1.558** & 1.572** & & 12.004** & 12.599* & & 31.103** & 33.614* & & 91.264* & 92.473* \\MSMD(8) MLE&
1.562** & 1.557** & & 12.135** & 12.083** & & 31.678** & 31.452** & & 90.499* & 90.801* \\MSMD(6) Bin-Whittle&
1.594** & 1.635  & & 12.972* & 13.260  & & 35.308** & 35.093* & & 98.206* & 95.891* \\MSMD(8) Bin-Whittle&
1.594** & 1.637  & & 12.985* & 13.274  & & 35.357** & 35.113* & & 98.348* & 95.914* \\MSMD(6) Log-Whittle&
1.595** & 1.637  & & 13.020* & 13.275  & & 35.475** & 35.105* & & 98.655* & 95.888* \\MSMD(8) Log-Whittle&
1.595** & 1.638  & & 13.022* & 13.286  & & 35.484** & 35.123* & & 98.689* & 95.911* \\MSMD(6)+MSMD(8)+LMSD MLE&
1.556** & 1.562** & & 11.980** & 12.306** & & 31.162** & 31.779** & & 86.533** & 87.319** \\MSMD(6)+MSMD(8)+LMSD Bin-Whitt.&
1.571** & 1.573** & & 12.017** & 12.161** & & 33.276* & 32.811** & & 91.385** & 89.567** \\MSMD(6)+MSMD(8)+LMSD Log-Whitt.&
1.571** & 1.574** & & 11.987** & 12.195** & & 33.355*** & 32.812** & & 91.590** & 89.556** \\
\midrule	
\multicolumn{8}{l}{C. Japanese Yen} \\
ACD & 1.769 & 1.771 & & 14.624 & 14.658 & &
43.349 & 43.432 & & 133.613 & 133.654 \\LMSD &
1.714** & 1.793  & & 13.149** & 13.694* & & 36.577** & 37.600** & & 107.462** & 109.681** \\MSMD(6) MLE&
1.712** & 1.715* & & 12.984** & 12.685** & & 34.812** & 34.280** & & 103.209** & 102.523** \\MSMD(8) MLE&
1.719* & 1.730 & & 13.184** & 12.960** & & 35.509** & 34.777** & & 102.423** & 103.065** \\MSMD(6) Bin-Whittle&
1.756  & 1.829\Cross & & 13.947* & 14.254  & & 40.057** & 39.714** & & 119.285** & 114.268** \\MSMD(8) Bin-Whittle&
1.757  & 1.829\Cross & & 13.968* & 14.246  & & 40.119** & 39.675** & & 119.399** & 114.092** \\MSMD(6) Log-Whittle&
1.760  & 1.838\Cross & & 14.042 & 14.317  & & 40.378** & 39.815** & & 120.163** & 114.273** \\MSMD(8) Log-Whittle&
1.760  & 1.835\Cross & & 14.033 & 14.288  & & 40.336** & 39.736** & & 119.983** & 114.055** \\MSMD(6)+MSMD(8)+LMSD MLE&
1.706** & 1.735  & & 12.883** & 12.988** & & 34.832** & 34.727** & & 100.499** & 100.807** \\MSMD(6)+MSMD(8)+LMSD Bin-Whitt.&
1.720** & 1.745  & & 12.970** & 13.176** & & 37.390** & 36.752** & & 111.419** & 107.634** \\MSMD(6)+MSMD(8)+LMSD Log-Whitt.&
1.722** & 1.749  & & 12.934** & 12.950** & & 37.533** & 36.779** & & 111.809** & 107.578** \\
\bottomrule	
\end{tabular}		
\caption{Mean square error (MSE) of out-of-sample forecasts for (A) Swiss franc, (B) Euro and (C) Japanese Yen futures price durations for 1, 5, 10 and 20 step ahead horizons. We use * and ** to denote that a competing model has significantly lower MSE (is better) in comparison to the ACD model at 95\% and 99\% significance levels, respectively. We use \Cross and \Cross\Cross to denote that a competing model has significantly higher MSE (is worse) in comparison to the ACD model at 95\% and 99\% significance levels, respectively.
 \label{tab:forecastsMSE}}
\end{table}	
\end{landscape}

\clearpage

\begin{landscape}
\begin{table}
\footnotesize
\centering				
\begin{tabular}{llllllllllll}	
\toprule							
 & \multicolumn{2}{c}{$h=1$} & &   \multicolumn{2}{c}{$h=5$} & &  \multicolumn{2}{c}{$h=10$} & &  \multicolumn{2}{c}{$h=20$}	  \\
\cmidrule{2-3}\cmidrule{5-6}\cmidrule{8-9}\cmidrule{11-12}				
 & Exp & $\mathcal{W(\kappa)}$ & & Exp & $\mathcal{W(\kappa)}$ & & Exp & $\mathcal{W(\kappa)}$ & & Exp & $\mathcal{W(\kappa)}$ \\
\midrule	
\multicolumn{8}{l}{A. Swiss franc} \\
ACD & 0.752 & 0.741 & & 2.411 & 2.227 & &
4.469 & 3.843 & & 8.976 & 6.824 \\LMSD &
0.770\Cross \Cross & 0.784\Cross \Cross & & 2.249** & 2.261  & & 3.870** & 3.869  & & 6.862** & 6.791  \\MSMD(6) MLE&
0.742* & 0.724** & & 2.153** & 2.084** & & 3.608** & 3.488** & & 6.730** & 6.567  \\MSMD(8) MLE&
0.741* & 0.738  & & 2.148** & 2.095** & & 3.600** & 3.496** & & 6.628** & 6.578  \\MSMD(6) Bin-Whittle&
0.741* & 0.750\Cross & & 2.226** & 2.270  & & 3.789** & 3.828  & & 6.747** & 6.767  \\MSMD(8) Bin-Whittle&
0.741* & 0.751\Cross & & 2.227** & 2.271 & & 3.791** & 3.830  & & 6.751** & 6.768  \\MSMD(6) Log-Whittle&
0.741* & 0.751\Cross & & 2.229** & 2.275 & & 3.795** & 3.834  & & 6.756** & 6.771  \\MSMD(8) Log-Whittle&
0.741* & 0.751\Cross & & 2.229** & 2.275 & & 3.796** & 3.834  & & 6.758** & 6.771  \\MSMD(6)+MSMD(8)+LMSD MLE&
0.748  & 0.746  & & 2.177** & 2.134** & & 3.653** & 3.541** & & 6.605** & 6.463** \\MSMD(6)+MSMD(8)+LMSD Bin-Whitt.&
0.746  & 0.753  & & 2.174** & 2.155** & & 3.750** & 3.743** & & 6.677** & 6.640** \\MSMD(6)+MSMD(8)+LMSD Log-Whitt.&
0.747  & 0.753  & & 2.161** & 2.117** & & 3.753** & 3.746** & & 6.681** & 6.642** \\
\midrule	
\multicolumn{8}{l}{B. Euro} \\
ACD & 0.789 & 0.788 & & 2.515 & 2.513 & &
4.303 & 4.299 & & 7.462 & 7.458 \\LMSD &
0.795  & 0.810\Cross \Cross & & 2.451 & 2.474  & & 4.090* & 4.131 & & 6.808** & 6.861** \\MSMD(6) MLE&
0.777** & 0.784  & & 2.381** & 2.455  & & 3.986** & 4.170  & & 7.096  & 7.193  \\MSMD(8) MLE&
0.777** & 0.780 & & 2.390** & 2.390** & & 3.984** & 3.996** & & 7.029 & 7.048  \\MSMD(6) Bin-Whittle&
0.779* & 0.790  & & 2.506  & 2.545  & & 4.265  & 4.280  & & 7.363  & 7.327  \\MSMD(8) Bin-Whittle&
0.779* & 0.790  & & 2.507  & 2.547  & & 4.267  & 4.281  & & 7.367  & 7.328  \\MSMD(6) Log-Whittle&
0.780* & 0.790  & & 2.509  & 2.547  & & 4.272  & 4.281  & & 7.376  & 7.327  \\MSMD(8) Log-Whittle&
0.780* & 0.790  & & 2.509  & 2.548  & & 4.272  & 4.282  & & 7.377  & 7.328  \\MSMD(6)+MSMD(8)+LMSD MLE&
0.780* & 0.789  & & 2.401** & 2.442* & & 3.980** & 4.050* & & 6.872* & 6.930* \\MSMD(6)+MSMD(8)+LMSD Bin-Whitt.&
0.779* & 0.785  & & 2.401* & 2.422 & & 4.134** & 4.131* & & 7.057** & 7.018* \\MSMD(6)+MSMD(8)+LMSD Log-Whitt.&
0.780* & 0.785  & & 2.392* & 2.421 & & 4.137** & 4.132* & & 7.063** & 7.019* \\
\midrule	
\multicolumn{8}{l}{C. Japanese Yen} \\
ACD & 0.855 & 0.855 & & 2.801 & 2.804 & &
4.689 & 4.694 & & 8.348 & 8.349 \\LMSD &
0.880\Cross \Cross & 0.908\Cross \Cross & & 2.723 & 2.776  & & 4.384** & 4.451* & & 7.581** & 7.688** \\MSMD(6) MLE&
0.844* & 0.840* & & 2.631** & 2.598** & & 4.207** & 4.164** & & 7.409** & 7.389** \\MSMD(8) MLE&
0.846* & 0.845 & & 2.648** & 2.624** & & 4.226** & 4.204** & & 7.378** & 7.415** \\MSMD(6) Bin-Whittle&
0.855  & 0.874\Cross & & 2.744* & 2.772  & & 4.530** & 4.513* & & 7.976** & 7.819** \\MSMD(8) Bin-Whittle&
0.855  & 0.874\Cross & & 2.746* & 2.771  & & 4.532** & 4.511* & & 7.979** & 7.813** \\MSMD(6) Log-Whittle&
0.856  & 0.876\Cross \Cross & & 2.751 & 2.777  & & 4.543** & 4.517* & & 8.001** & 7.819** \\MSMD(8) Log-Whittle&
0.856  & 0.875\Cross \Cross & & 2.750 & 2.775  & & 4.541** & 4.513* & & 7.996** & 7.812** \\MSMD(6)+MSMD(8)+LMSD MLE&
0.853  & 0.861  & & 2.656** & 2.667** & & 4.229** & 4.230** & & 7.328** & 7.349** \\MSMD(6)+MSMD(8)+LMSD Bin-Whitt.&
0.857  & 0.869\Cross & & 2.661** & 2.684** & & 4.382** & 4.349** & & 7.677** & 7.573** \\MSMD(6)+MSMD(8)+LMSD Log-Whitt.&
0.858  & 0.870\Cross & & 2.646** & 2.649** & & 4.388** & 4.350** & & 7.689** & 7.572** \\
\bottomrule	
\end{tabular}		
\caption{Mean absolute deviations (MAD) of out-of-sample forecasts for (A) Swiss franc, (B) Euro and (C) Japanese Yen futures price durations for 1, 5, 10 and 20 step ahead horizons. We use * and ** to denote that a competing model has significantly lower MAD (is better) in comparison to the ACD model at 95\% and 99\% significance levels, respectively. We use \Cross and \Cross\Cross to denote that a competing model has significantly higher MAD (is worse) in comparison to the ACD model at 95\% and 99\% significance levels, respectively.\label{tab:forecastsMAD}}
\end{table}	
\end{landscape}

\clearpage


\end{singlespace}

\end{document}